\documentclass[aps,prd,showpacs,floatfix,nofootinbib,twocolumn,10pt]{revtex4}

\usepackage{color,amsmath,amssymb,graphicx,latexsym,subfigure}
\usepackage{threeparttable,multirow,txfonts}

\begin{document}

\title{Propagation of cosmic rays in the AMS-02 era}

\author{Qiang Yuan$^{a,b}$\footnote{Corresponding author: yuanq@pmo.ac.cn}}
\author{Su-Jie Lin$^c$}
\author{Kun Fang$^c$}
\author{Xiao-Jun Bi$^c$\footnote{Corresponding author: bixj@ihep.ac.cn}}

\affiliation{
$^a$Key Laboratory of Dark Matter and Space Astronomy, Purple Mountain 
Observatory, Chinese Academy of Sciences, Nanjing 210008, P.R.China \\
$^b$School of Astronomy and Space Science, University of Science and 
Technology of China, Hefei 230026, P.R.China\\
$^c$Key Laboratory of Particle Astrophysics, Institute of High Energy 
Physics, Chinese Academy of Sciences, Beijing 100049, P.R.China
}

\begin{abstract}

In this work we use the newly reported Boron-to-Carbon ratio (B/C) from 
AMS-02 and the time-dependent proton fluxes from PAMELA and AMS-02 to 
constrain the source and propagation parameters of cosmic rays in the 
Milky Way. A linear correlation of the solar modulation parameter with 
solar activities is assumed to account for the time-varying cosmic ray 
fluxes. A comprehensive set of propagation models, with/without 
reacceleration or convection, have been discussed and compared. We find 
that only the models with reacceleration can self-consistently fit both 
the proton and B/C data. The rigidity dependence slope of the diffusion 
coefficient, $\delta$, is found to be about $0.38-0.50$ for the  
diffusion-reacceleration models. The plain diffusion and diffusion-convection 
models fit the data poorly. We compare different model predictions of 
the positron and antiproton fluxes with the data. We find that the 
diffusion-reacceleration models over-produce low energy positrons, 
while non-reacceleration models give better fit to the data. As for 
antiprotons, reacceleration models tend to under-predict low energy 
antiproton fluxes, unless a phenomenological modification of the 
velocity-dependence of the diffusion coefficient is applied. Our results 
suggest that there could be important differences of the propagation for
nuclei and leptons, in either the Milky Way or the solar heliosphere.

\end{abstract}

\date{\today}

%95.35.+d: Dark matter
%96.50.S-: Cosmic rays
\pacs{95.35.+d,96.50.S-}

\maketitle

\section{Introduction}

The propagation of cosmic rays (CRs) in the Milky Way is a fundamental
question to understand the origin and interactions of Galactic CRs.
It also provides us a useful tool to probe the properties of the
interstellar medium (ISM). It is well known that the charged CRs will
propagate diffusively in the Galactic magnetic field, experiencing
possibly the reacceleration, convection, spallation and energy loss
processes \cite{1990cup..book.....G,2007ARNPS..57..285S}. The
propagation process can be described with the diffusive transport
equation \cite{1990acr..book.....B,1990cup..book.....G}. Depending
on different simplifications, the transport equation can be solved
analytically \cite{1992ApJ...390...96W,1993A&A...267..372B,
2001ApJ...555..585M,2002A&A...394.1039M,2004ApJ...612..238S}.
Also there were efforts to include most of the relevant processes
and the observation-based astrophysical inputs, and to solve the
propagation equation numerically, e.g., GALPROP
\cite{1998ApJ...509..212S,1998ApJ...493..694M} and DRAGON
\cite{2008JCAP...10..018E}.

To understand the propagation of CRs is not only important for the
CR physics itself, but also the basis of searching for the exotic
signal from particle dark matter. The propagation of CRs couples
closely with the production, leading to the entanglement between 
source parameters and propagation parameters. Fortunately, the 
spallation of the CR nuclei when colliding with the ISM will produce 
secondary nuclei (with kinetic energy per nucleon unchanged). 
The ratio between those secondary nuclei and the parent nuclei will 
cancel out the source information, leaving basically the propagation 
effect. Widely used are the Boron-to-Carbon (B/C) and sub-Iron-to-Iron 
((Sc+Ti+V)/Fe) ratios. The unstable-to-stable ratio of the secondary 
isotopes plays another important role to constrain the CR propagation. 
The unstable nuclei with lifetimes comparable to the diffusion time 
of the CRs, such as $^{10}$Be ($\tau=1.39\times10^6$ yr) and $^{26}$Al 
($\tau=7.17\times10^5$ yr), can be used as the clocks to measure the 
residual time of CRs in the Milky Way halo.

Many works have been dedicated to using the secondary-to-primary ratios
and the unstable-to-stable isotope ratios to constrain the CR propagation
parameters (see e.g., \cite{1990ApJ...349..625S,1991ApJ...374..356M,
1998ApJ...509..212S,2001ApJ...563..768Y,2001ApJ...555..585M,
2005JCAP...09..010L,2009ApJ...697..106A,2010JCAP...06..022P,
2010APh....34..274D,2012ApJ...752...69O}). However, due to the large
number of the model parameters and the degeneracy between different
parameters, the investigation of the parameter space is incomplete
and the conclusion might be biased. In addition, more and more data
have been accumulated nowadays. It is necessary to combine different
data sets in a statistical way. Recently several works employed the
Markov Chain Monte Carlo (MCMC) method to try to take a full scan of
the parameter space with large samples of the data
\cite{2009A&A...497..991P,2010A&A...516A..66P,2011ApJ...729..106T,
2015JCAP...09..049J,2016ApJ...824...16J,2016PhRvD..94l3019K,
2016PhRvD..94l3007F}. 
The MCMC method is known to be efficient for the minimization of
high-dimensional problem and is widely used in different areas.

We have developed a tool, {\tt CosRayMC}, through embeding the CR
propagation code in the MCMC sampler \cite{2002PhRvD..66j3511L},
which have already been applied to the study of the CR lepton excesses 
\cite{2010PhRvD..81b3516L,2012IJMPA..2750024L,2012PhRvD..85d3507L,
2015APh....60....1Y,2013PhLB..727....1Y,2015JCAP...03..033Y}.
In light of the newly reported CR nuclei and B/C data by PAMELA and
AMS-02, we apply this tool to re-visit the CR propagation and constrain
the propagation parameters in this work. Compared with previous studies
\cite{2011ApJ...729..106T,2015JCAP...09..049J,2016ApJ...824...16J,
2016PhRvD..94l3019K,2016PhRvD..94l3007F}, we will present an extensive
study of different propagation models, including the plain diffusion
scenario, the diffusion reacceleration scenario and the diffusion
convection scenario. Furthermore, we will employ a phenomenological
treatment of the time-dependent solar modulation based on the solar
activities. Finally, the predicted positron and antiproton fluxes of
different propagation models will be compared with the data as a
consistency check.

This paper is organized as follows. In Sec. II we define the propagation
model configurations. In Sec. III we describe the fitting procedure.
The fitting results and expectations of secondary positron and antiproton
fluxes are presented in Sec. IV. We discuss our results in Sec. V, and
finally conclude in Sec. VI.

\section{Propagation models}

Galactic CRs are accelerated in cosmic accelerators such as supernova 
remnants and pulsars before they are injected into the ISM. During their
propagation in the Galaxy, secondary particles can be produced by the 
collisions between primary CRs and the ISM. The propagation of CRs 
in the Galaxy is usually described by the diffusive transport equation
\begin{equation*}
  \begin{split}
    \frac{\partial \psi}{\partial t} =& Q(\mathbf{x}, p) + \nabla \cdot \left( D_{xx}\nabla\psi - \mathbf{V}_{c}\psi \right)
    + \frac{\partial}{\partial p}p^2D_{pp}\frac{\partial}{\partial p}\frac{1}{p^2}\psi \\
    &- \frac{\partial}{\partial p}\left[ \dot{p}\psi - \frac{p}{3}\left( \nabla\cdot\mathbf{V}_c\psi \right) \right]
    - \frac{\psi}{\tau_f} - \frac{\psi}{\tau_r},
  \end{split}
  \label{propagation_equation}
\end{equation*}
where $\psi$ is the differential density of CR particles per momentum
interval, $Q$ is the source term, $D_{xx}$ is the spatial diffusion
coefficient, $\mathbf{V}_c$ is the convective velocity, $D_{pp}$ is
the diffusion coefficient in the momentum space describing the
reacceleration effect, $\dot{p}\equiv\mathrm{d}p/\mathrm{d}t$ is the
momentum loss rate, $\tau_f$ and $\tau_r$ are correspondingly the time
scales for nulear fragmentation and radioactive decay.

The diffusion coefficient is usually assumed to vary with rigidity by
a power-law form
\begin{equation}
  D_{xx}=D_0\beta^\eta\left( \frac{R}{R_0} \right)^\delta,
  \label{D_xx}
\end{equation}
where $D_0$ is the normalization factor, $R_0$ is a reference rigidity,
$\delta$ is the power-law index which depends on the property of
turbulence in the interstellar medium (ISM), $\beta$ is the velocity
in unit of light speed, and $\eta$ is a phenomenological parameter
describing the velocity dependence of the diffusion coefficient at low
energies, which is generally to be 1. For single power-law form of
$D_{xx}$, we fix $R_0$ to be 4 GV. For the broken power-law case
(see below), $R_0$ is left to be free in the fitting.

We assume the convection velocity linearly and continuously vary from
the Galactic disk to halo, $\mathbf{V}_c=\mathbf{z}\cdot
\mathrm{d}V_c/\mathrm{d}z$, where $\mathbf{z}$ is the position vector 
in the vertical direction to the Galactic disk. Such a form can
avoid the discontinuity at the Galactic plane.

The reacceleration effect would lead to a diffusion in the momentum space. 
Its diffusion coefficient in momentum space, $D_{pp}$, is related with the 
spatial diffusion coefficient as \cite{1994ApJ...431..705S}
\begin{equation}
  D_{pp}D_{xx}=\frac{4p^2v_A^2}{3\delta(4-\delta^2)(4-\delta)\omega},
  \label{equation_Dpp}
\end{equation}
where the $v_A$ is the Alfven velocity and $\omega$ is the ratio of
magnetohydrodynamic wave energy density to magnetic field energy
density. Since $\omega$ can be effectively absorbed in $v_A$, we assume 
it to be 1.

The source function $Q(\mathbf{x},p)$ is expressed as $f(\mathbf{x})q(p)$, 
where $f(\mathbf{x})$ is the spatial distribution and $q(p)$ is the 
injection energy spectrum of CR sources. The spatial distribution is
assumed to follow that of supernova remnants
\begin{equation}
  f(r,z) = \left(\frac{r}{r_\odot}\right)^{1.25} 
  {\rm exp}\left(-3.56\cdot\frac{r-r_\odot}{r_\odot}\right)
  {\rm exp}\left(-\frac{|z|}{z_s}\right)\,,
  \label{spatial_distribution}
\end{equation}
with parameters slightly adjusted to match the Galactic diffuse $\gamma$-ray
emission and the ratio of H$_2$ to CO \cite{2007ARNPS..57..285S,
2011ApJ...729..106T}; $r_\odot=8.5$ kpc is the distance from the Sun to the
Galactic center; $z_s\approx 0.2$ kpc is the characteristic height of the
Galactic disk. The nuclei injection spectrum is assumed to be a broken 
power-law function of rigidity
\begin{equation}
  q(R)\propto\left\{
    \begin{array}{ll}
      \left( R / R_{\mathrm{br}}\right)^{-\nu_1}, & R < R_{\mathrm{br}} \\
      \left( R / R_{\mathrm{br}} \right)^{-\nu_2}, &  R \ge R_{\mathrm{br}}
    \end{array}.\right.
  \label{injection_equation}
\end{equation}
Power-law form of particle spectrum is expected from the simple shock
acceleration mechanism. However, it has been found that single power-law
spectrum is somehow not enough to describe the observational data,
especially when there is strong reacceleration of CRs 
\cite{2011ApJ...729..106T}. The observations of $\gamma$-ray emission
from a few supernova remnants which are interacting with molecular
clouds also suggest a broken-power law of CRs in/around the source
\cite{2013Sci...339..807A}. Note that we neglect the potential second 
break at hundreds of GV of the CR nuclei \cite{2007BRASP..71..494P,
2010ApJ...714L..89A,2011Sci...332...69A,2015PhRvL.114q1103A,
2015PhRvL.115u1101A}, which is beyond the energy range we are interested 
in. Since we focus on the B/C ratio, the small difference between the
spectra of protons and heavier nuclei \cite{2011Sci...332...69A,
2015PhRvL.115u1101A} is also neglected.

The diffusive nature of charged particles in the Milky Way has been
well established \cite{1990cup..book.....G}. However, whether the
reacceleration and/or convection plays significant roles in regulating
the propagation of CRs is unclear. The widely existed galactic winds
suggest that convective transport of CRs may be relevant 
\cite{1976ApJ...208..900J}. On the other hand, the observed peak of 
the B/C around $\sim1$ GeV/n by HEAO-3 \cite{1990A&A...233...96E}
may require an effective reacceleration \cite{2002ApJ...565..280M}.
While the reacceleration model can fit the B/C data, it would 
under-predict antiprotons \cite{2002ApJ...565..280M}. An adjustment
of the $\eta$ parameter in the diffusion coefficient was introduced
to solve such a discrepancy \cite{2010APh....34..274D}. 
The modification of the low energy diffusion coefficient is
also physically motivated from the potential resonant interaction
of CR particles and the magnetohydrodynamic (MHD) waves which results 
in dissipation of such waves \cite{2006ApJ...642..902P}.

In this work we will test all these kinds of models with the new
observational data. Specifically, the propagation models include: 
1) the plain diffusion (PD) model without reacceleration and convection,
2) the diffusion convection (DC) model, 3) the diffusion convection model 
with a break of the rigidity-dependence of the diffusion coefficient 
(with $\delta=0$ below the break rigidity $R_0$ \cite{2002ApJ...565..280M}; 
DC2), 4) the diffusion reacceleration (DR) model, 5) the diffusion 
reacceleration model with $\eta$ left free to fit (DR2), and 6) the 
diffusion reacceleration convection (DRC) model. The relevant propagation
parameters are $(D_0,\,\delta,\,z_h,\,v_A,\,dV_c/dz,\,R_0,\,\eta)$.

We keep in mind that the above described propagation framework is
actually simplfied. The diffusion coefficient may vary in the Milky Way
due to different magnetic field distributions in the disk and halo
(e.g., \cite{2012ApJ...752L..13T,2016ApJ...819...54G}). In particular,
CRs may be confined much longer around the sources than expected due to 
non-linear self-generation of MHD waves via streaming instability
\cite{2016PhRvD..94h3003D}. These complications are less clear and beyond 
the scope of the current work. The caveat is that considering these 
effects may result in different results from the adopted framework
(see e.g., \cite{2016PhRvD..94l3007F}).

\section{Fitting procedure}

\subsection{\tt CosRayMC}

The {\tt CosRayMC} code is a combination of the numerical propagation
code {\tt GALPROP}\footnote{http://galprop.stanford.edu/}
\cite{1998ApJ...509..212S,1998ApJ...493..694M} and the MCMC sampler
(adapted from CosmoMC \cite{2002PhRvD..66j3511L}). The MCMC technique
is widely applied in astrophysics and cosmology to investigate the
high-dimensional parameter space from observational data. It works
in the Bayesian framework. The posterior probability of model
parameters $\boldsymbol{\theta}$ in light of the observational data
$D$ is ${\mathcal P}(\boldsymbol{\theta}|D)\propto{\mathcal P}
(D|\boldsymbol{\theta}){\mathcal P}(\boldsymbol{\theta})$, where
${\mathcal P}(D|\boldsymbol{\theta})$ is the likelihood and
${\mathcal P}(\boldsymbol{\theta})$ is the prior probability of
$\boldsymbol{\theta}$.

The Markov chain is generated following the Metropolis-Hastings
algorithm. The general process is as follows. One first proposes a
random step in the parameter space. Then the acceptance probability
is calculated by the ratio of the target probabilities of this
proposed point to the former one. If the proposed point is accepted,
then repeat this procedure. Otherwise, go back to the former point
and have another trial. The stationary distribution of the chain
samples will approaches the target probability distribution
${\mathcal P}(\boldsymbol{\theta}|D)$. For more details, one can
refer to \cite{Neal1993,Gamerman1997}.

\subsection{Data sets}

We adopt the most recently available accurate data sets of CRs by PAMELA
and AMS-02 in our fittings. For the B/C ratio, we employ the just-released
data by AMS-02 which cover an energy range of hundreds of MeV/n to TeV/n
\cite{2016PhRvL.117w1102A}. In order to have better constraints on the
low energy behavior of the B/C ratio, we also employ the data from
ACE-CRIS\footnote{http://www.srl.caltech.edu/ACE/ASC/level2/lvl2DATA\_CRIS.html}
with the same period as that of AMS-02. To constrain the lifetime of CRs in
the Galaxy, we also use the $^{10}$Be/$^9$Be data from some old measurements:
Ulysses \cite{1998ApJ...501L..59C}, ACE \cite{2001ApJ...563..768Y},
Voyager \cite{1999ICRC....3...41L}, IMP \cite{1988SSRv...46..205S},
ISEE-3 \cite{1988SSRv...46..205S}, and ISOMAX \cite{2004ApJ...611..892H}.
The proton fluxes are employed to constrain the injection parameters
of CRs. As will be discussed in the next subsection, we will try to
give a more reasonable treatment of the solar modulation effect,
the time-dependent proton fluxes from 2006 to 2009 measured by PAMELA
\cite{2013ApJ...765...91A} and the average flux from 2011 to 2013 by
AMS-02 \cite{2015PhRvL.114q1103A} are used. Table \ref{table:ssn}
summarizes the observational time of each data sets.

\begin{table}[!htb]
\caption {Data taking time of various measurements and the average
modelled sunspot numbers one year before the data taking time.}
\begin{tabular}{lcccccccc}
\hline \hline
 & Time & $\bar{N}$ \\
\hline
%ACE(B/C) & 08/1997-04/1998 & 10.0 \\
ACE($^{10}$Be/$^9$Be) & 08/1997-04/1999 & 23.5 \\
ACE(B/C) & 05/2011-05/2016 & 54.3 \\
PAMELA-2006($p$) & 11/2006 & 17.4 \\
PAMELA-2007($p$) & 12/2007 & 7.3 \\
PAMELA-2008($p$) & 12/2008 & 3.0 \\
PAMELA-2009($p$) & 12/2009 & 1.0 \\
AMS-02($p$) & 05/2011-11/2013 & 40.8 \\
AMS-02(B/C) & 05/2011-05/2016 & 54.3 \\
\hline
PAMELA($\bar{p}$) & 07/2006-12/2008 & 10.0 \\
AMS-02($e^+$) & 05/2011-11/2013 & 40.8 \\
AMS-02($\bar{p}$) & 05/2011-05/2015 & 51.5 \\
\hline
\hline
\end{tabular}
\label{table:ssn}\\
Notes: the data below the middle line are not fitted.
\end{table}

\subsection{Solar modulation}

\begin{figure}[!htb]
\includegraphics[width=0.45\textwidth]{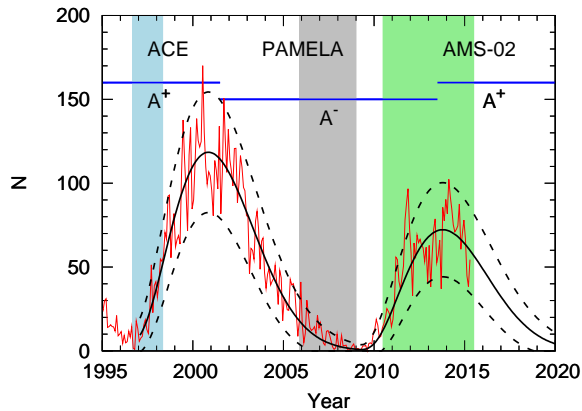}
\caption{Evolution of the sunspot numbers with time. The solid and dashed
lines show the predicted sunspot numbers and the $95\%$ intervals according
to the monitored data \cite{1999JGR...10422375H}. Shaded regions show
the periods of data taking (shifted leftwards by one year considering the
possible delay of modulation effect compared with the solar activity) by
ACE (for $^{10}$Be/$^9$Be), PAMELA (for protons) and AMS-02 (for all species)
detectors.
\label{fig:ssn}}
\end{figure}

In this work we use the force-field approximation to account for the solar
modulation of low energy CRs when propagating in the heliosphere
\cite{1968ApJ...154.1011G}. However, since the various data sets in
our work cover a wide time window in which solar activities varies much,
they should not share a common modulation potential. Fig. \ref{fig:ssn} 
shows the sunspot numbers of different time from 1995 to
present\footnote{https://solarscience.msfc.nasa.gov/SunspotCycle.shtml}.
The data we use are basically from the end of solar cycle 23 to the 
begining of solar cycle 24, except for the $^{10}$Be/$^9$Be data. 
More importantly they are roughly in the period that the polarity
of the solar magnetic field is in the same $A^-$ cycle. This enables
us to have a relatively simple approach of the solar modulation with
a correlation with solar activities.

Here we employ a linear evolution behavior of the modulation potential
with respect to the evolution of the sunspot number
\begin{equation}
\Phi=\Phi_0+\Phi_1\times\frac{N(t)}{N_{\rm max}},
\end{equation}
where $N_{\rm max}\approx72.2$ is the model predicted maximum sunspot
number in solar cycle 24 (shown by the solid line in Fig. \ref{fig:ssn};
\cite{1999JGR...10422375H}), $N(t)$ is the sunspot number during
which the data were collected, $\Phi_0$ and $\Phi_1$ are free parameters
which will be derived through fitting to the CR data. The average sunspot
numbers for various CR data taking time are given in Table \ref{table:ssn}.
Note that we always count the sunspot number for the time {\it one year}
before the actual data taking time, due to the possible delay of the
modulation effect compared with solar activity. This treatment is
consistent with the fact that the PAMELA proton flux in 12/2009 is
higher than that in 12/2008, while the solar minimum of cycle 23 ended
at the beginning of 2009. Given a typical speed of $\sim500$ km/s, solar 
winds need about one year to fill the heliosphere with a scale of $\sim$100
astronomical units, which further supports our treatment.

\section{Results}

\subsection{Fitting results of various models}

\begin{figure*}[!htb]
\includegraphics[width=0.7\textwidth]{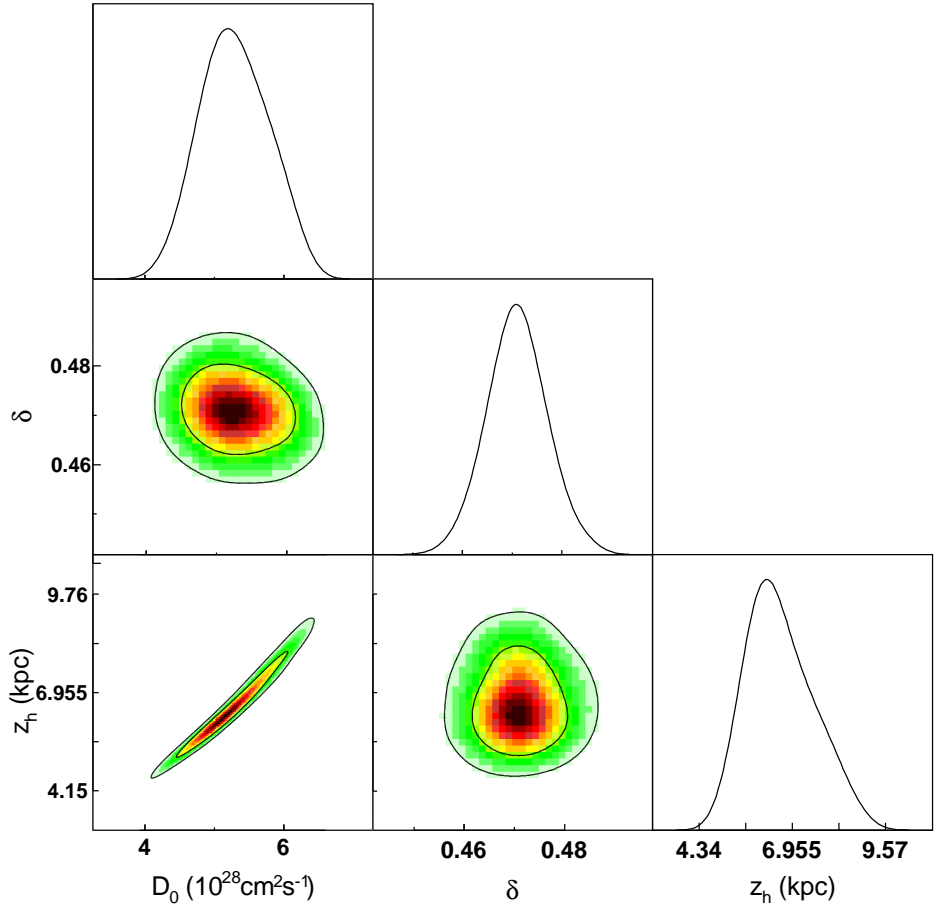}
\caption{Fitting 1-d probability distributions and 2-d credible regions
($68\%$ and $95\%$ credible levels from inside to outside) of the model
parameters in the PD scenario. 
\label{fig:tri_PD}}
\end{figure*}

\begin{figure*}[!htb]
\includegraphics[width=0.8\textwidth]{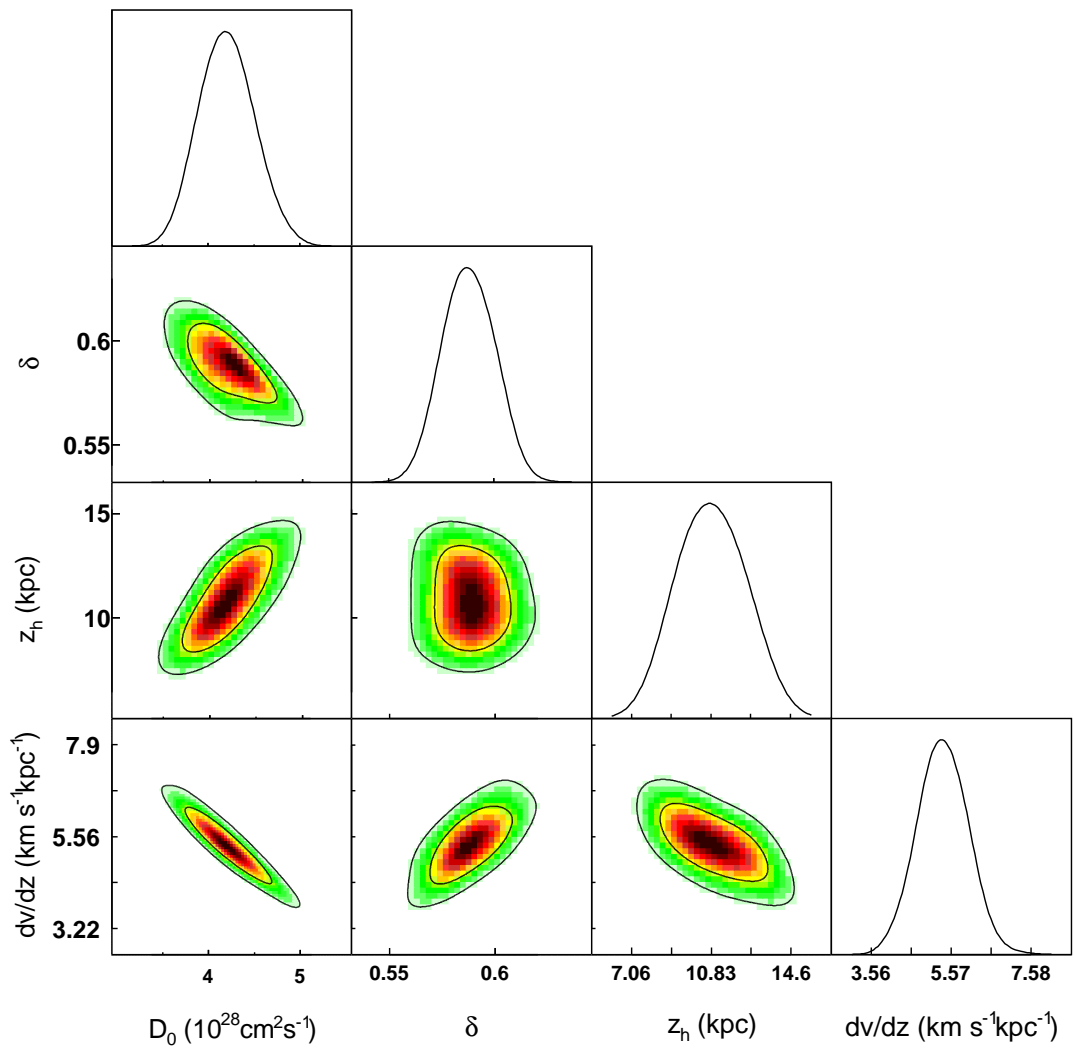}
\caption{Same as Fig. \ref{fig:tri_PD} but for the DC scenario (adding one
more parameter, $dV_c/dz$).
\label{fig:tri_DC}}
\end{figure*}

\begin{figure*}[!htb]
\includegraphics[width=0.8\textwidth]{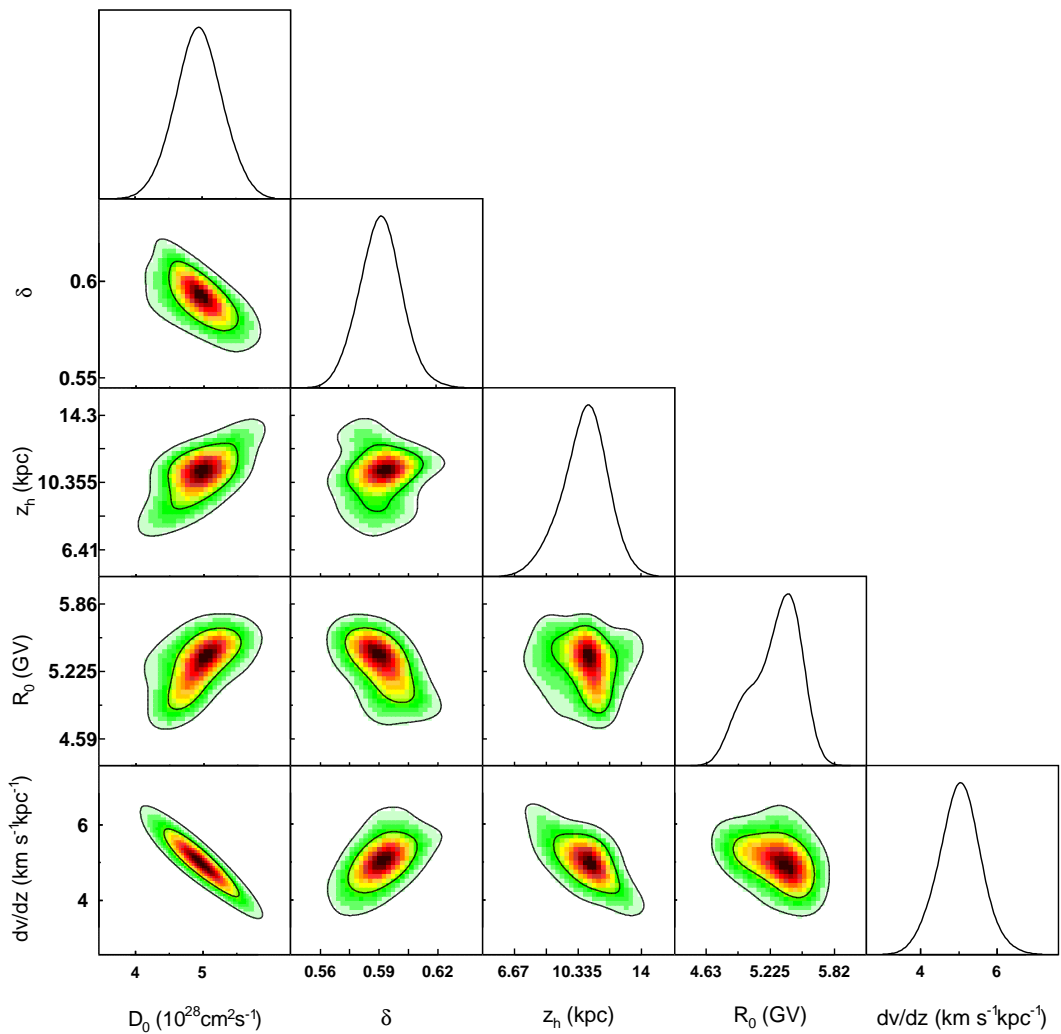}
\caption{Same as Fig. \ref{fig:tri_PD} but for the DC2 scenario (adding two
more parameters, $dV_c/dz$ and $R_0$).
\label{fig:tri_DC2}}
\end{figure*}

\begin{figure*}[!htb]
\includegraphics[width=0.8\textwidth]{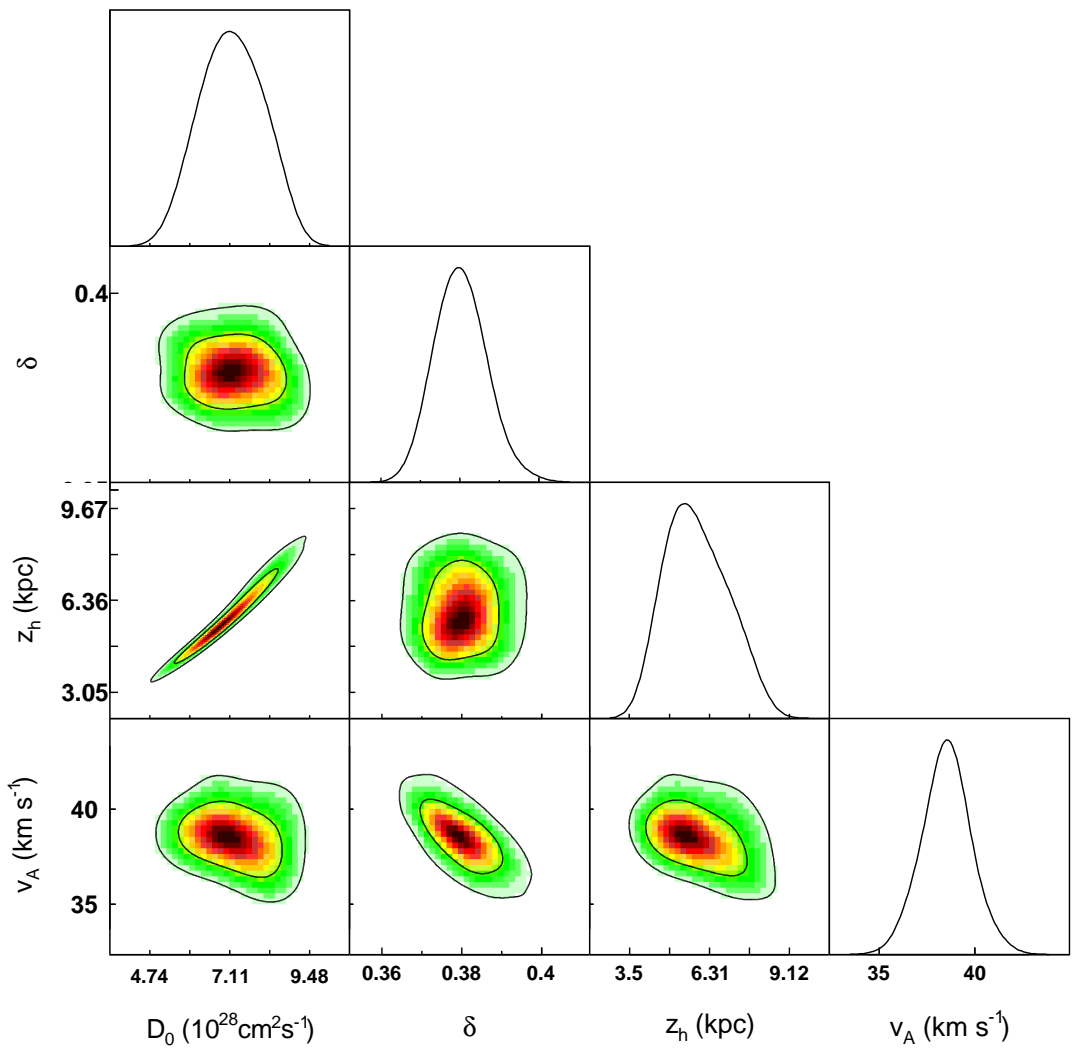}
\caption{Same as Fig. \ref{fig:tri_PD} but for the DR scenario (adding one
more parameter, $v_A$).
\label{fig:tri_DR}}
\end{figure*}

\begin{figure*}[!htb]
\includegraphics[width=0.8\textwidth]{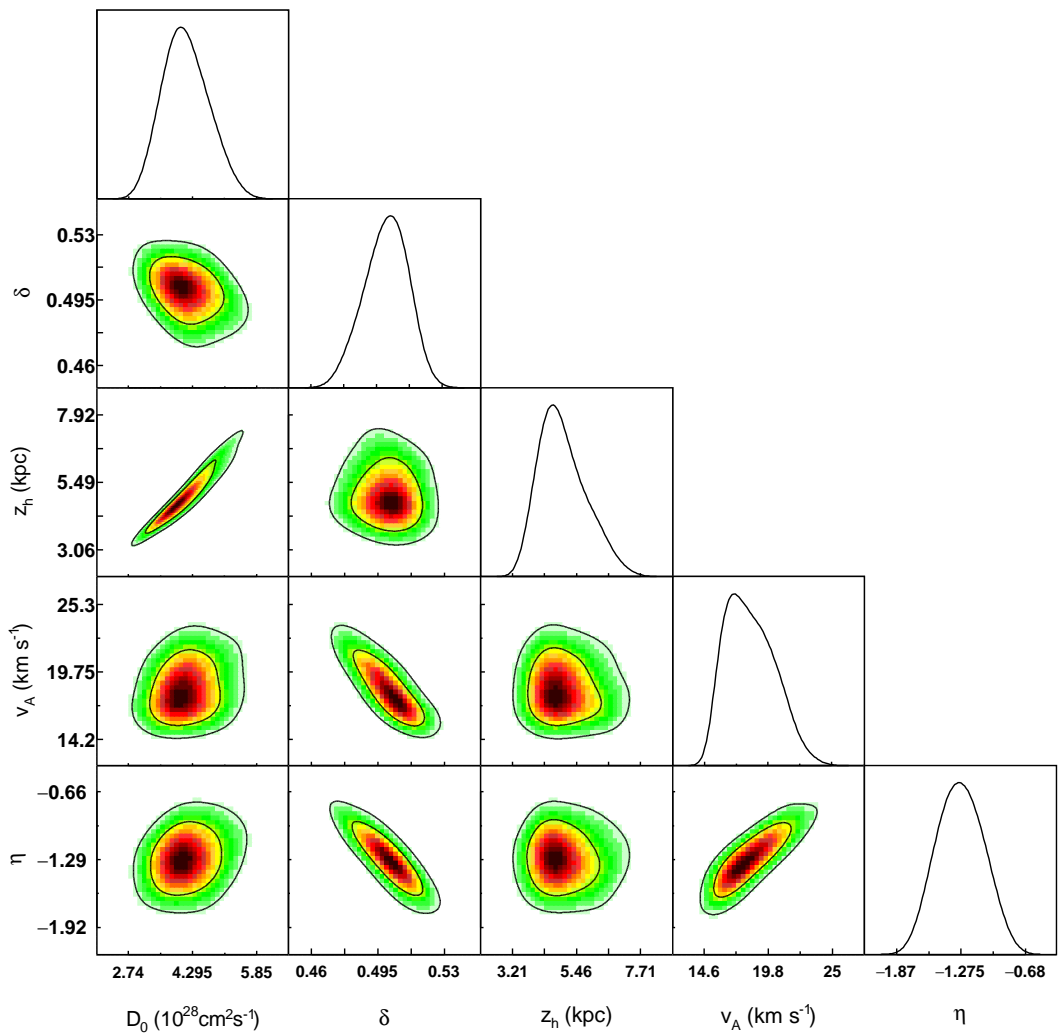}
\caption{Same as Fig. \ref{fig:tri_PD} but for the DR2 scenario (adding two
more parameters, $v_A$ and $\eta$).
\label{fig:tri_DR2}}
\end{figure*}

\begin{figure*}[!htb]
\includegraphics[width=0.8\textwidth]{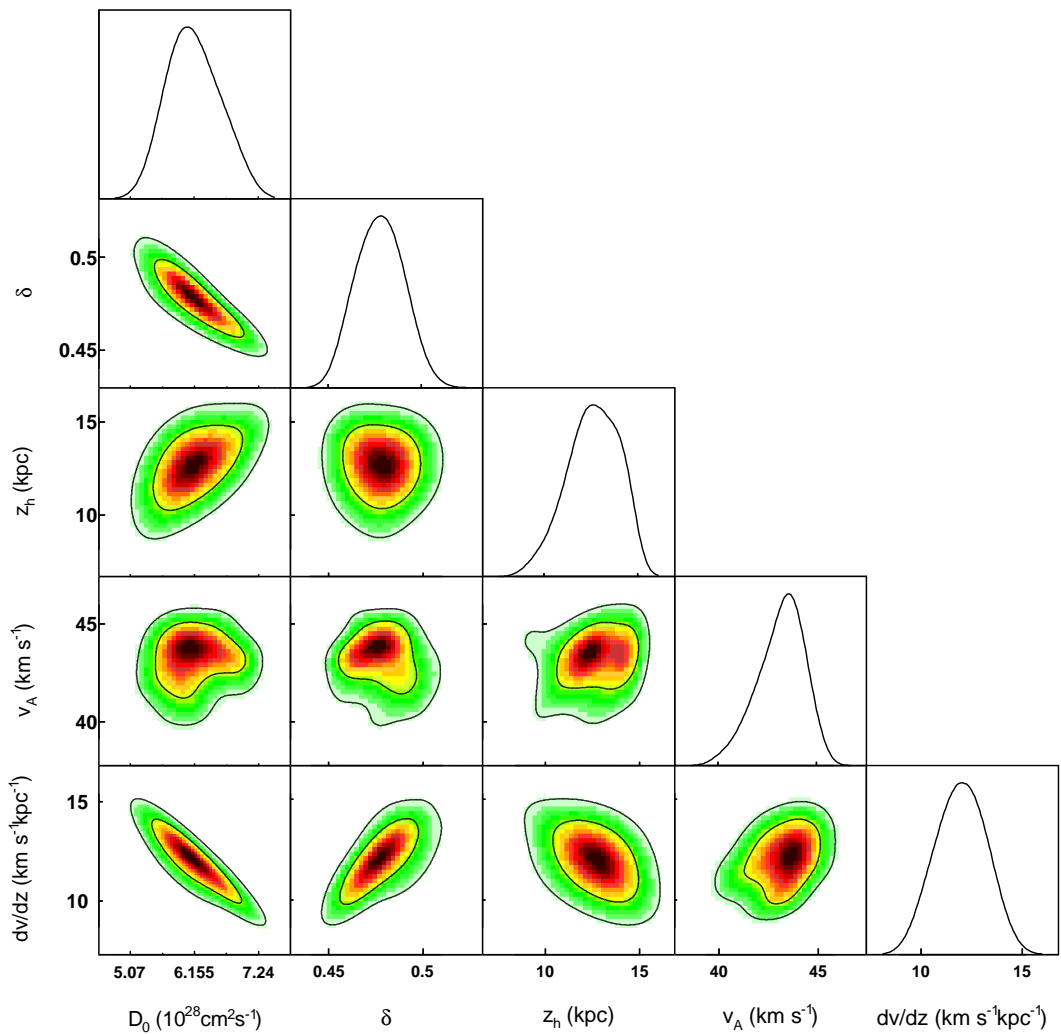}
\caption{Same as Fig. \ref{fig:tri_PD} but for the DRC scenario (adding two
more parameters, $v_A$ and $dV_c/dz$).
\label{fig:tri_DRC}}
\end{figure*}

\begin{figure*}[!htb]
\includegraphics[width=\textwidth]{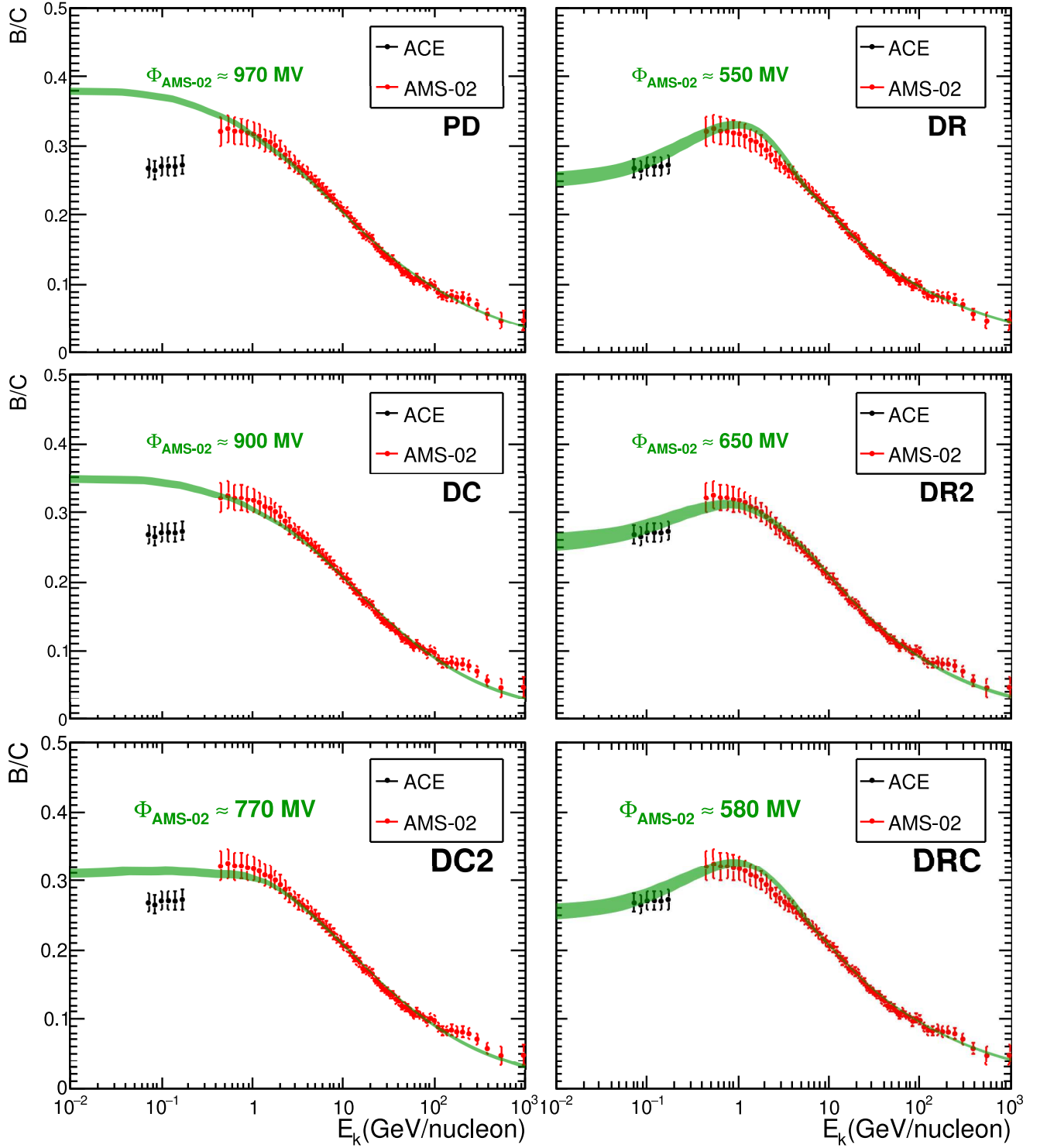}
\caption{$2\sigma$ bands of the B/C ratios for different PD propagation
models. The observational data are from: ACE \cite{2009ApJ...698.1666G} 
and AMS-02 \cite{2016PhRvL.117w1102A}.
\label{fig:bc}}
\end{figure*}

\begin{figure*}[!htb]
\includegraphics[width=\textwidth]{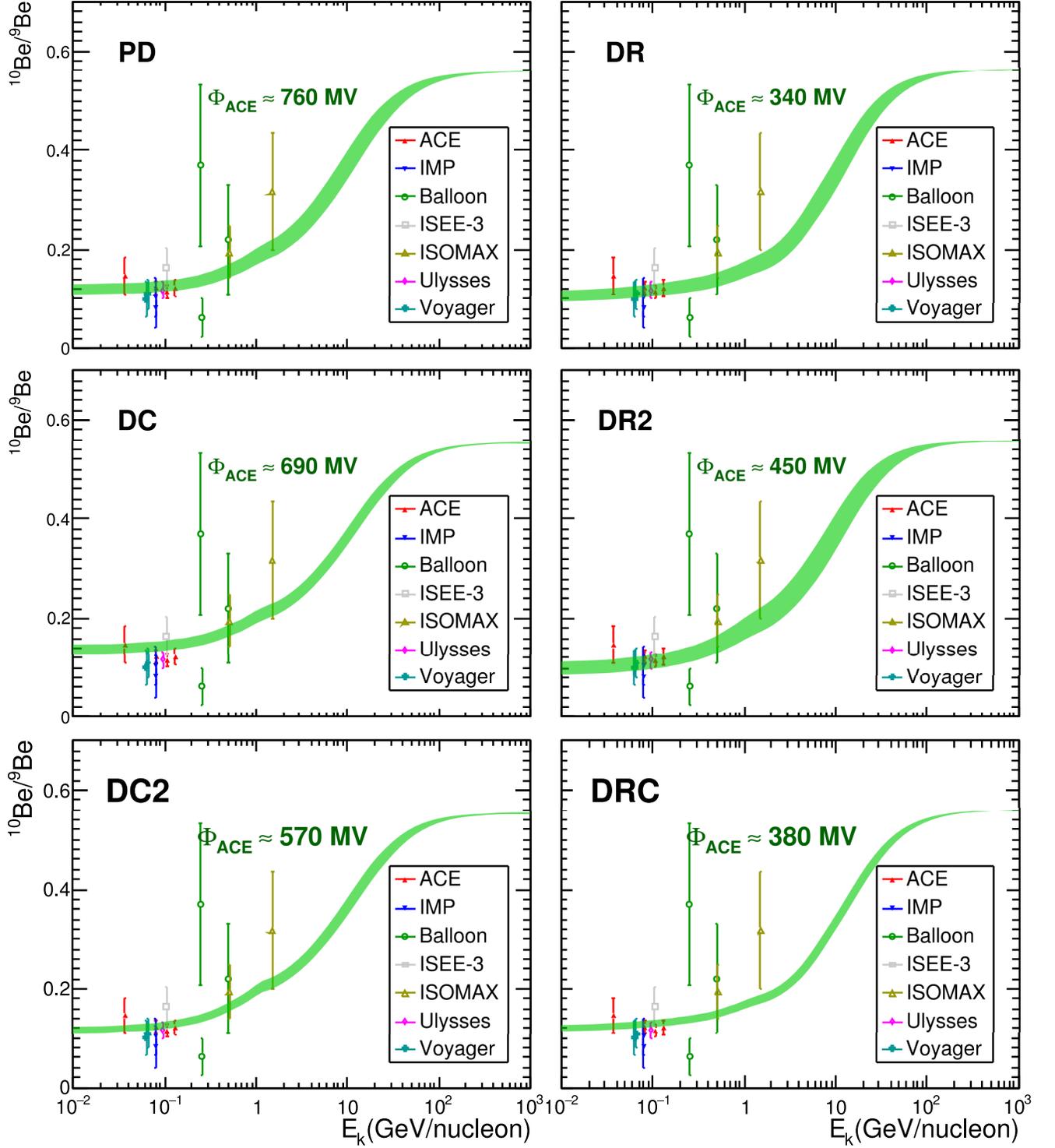}
\caption{$2\sigma$ bands of the $^{10}$Be/$^9$Be ratios for different
propagation models. The observational data are from:
Ulysses \cite{1998ApJ...501L..59C}, ACE \cite{2001ApJ...563..768Y},
Voyager \cite{1999ICRC....3...41L}, IMP \cite{1988SSRv...46..205S},
ISEE-3 \cite{1988SSRv...46..205S}, and ISOMAX \cite{2004ApJ...611..892H}.
\label{fig:be}}
\end{figure*}

\begin{figure*}[!htb]
\includegraphics[width=\textwidth]{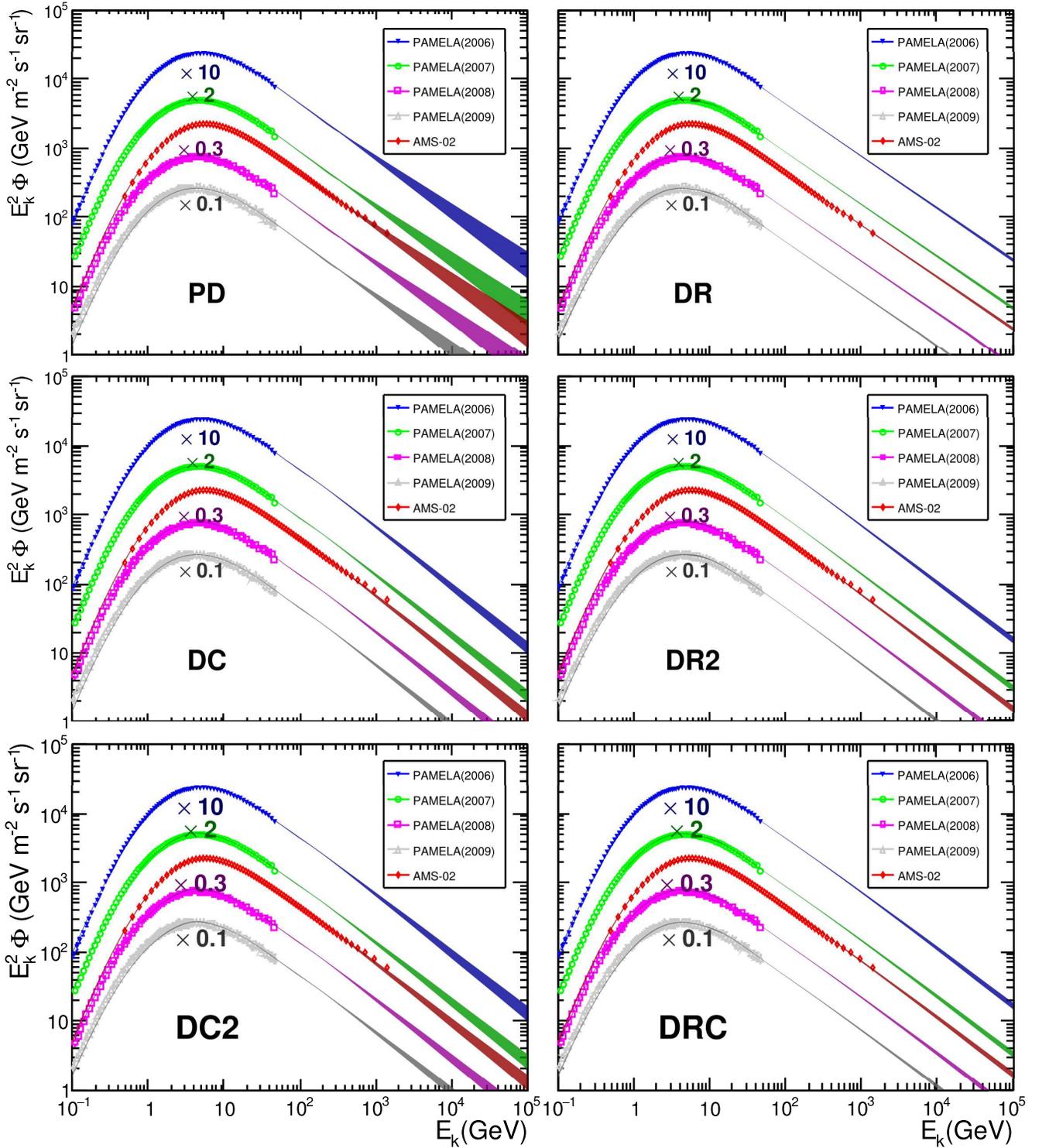}
\caption{Fitting $2\sigma$ bands of the proton spectra, compared to the
PAMELA results at four different epochs \cite{2013ApJ...765...91A}
and the AMS-02 data \cite{2015PhRvL.114q1103A}.
\label{fig:proton}}
\end{figure*}

We use the MCMC algorithm to determine the model parameters of the six
models as described in Sec. II through fitting to the data. The posterior 
mean and $68\%$ credible uncertainties of the model parameters are given 
in Table \ref{table:para}. Since the data are precise enough, we obtain
{\it statistically} good constraints on the model parameters. Some of the
model parameters, such as the injection spectral indices, are constrained
to a level of $\lesssim1\%$. The propagation parameters are constrained
to be about $10\%-20\%$, which are relatively large due to the degeneracy
among some of them. For the rigidity-dependence slope of the diffusion 
coefficient, $\delta$, the statistical error is only a few percent.
Compared with previous studies \cite{2011ApJ...729..106T,
2015JCAP...09..049J,2016ApJ...824...16J}, our results are widely improved. 
The one-dimensional (1-d) probability distributions and two-dimensional 
(2-d) confidence regions of the major propagation parameters are 
summarized in Figs. \ref{fig:tri_PD}-\ref{fig:tri_DRC}. We also show 
explicitly the comparison of the data with the fitting results 
(with $95\%$ credible bands) in Figs. \ref{fig:bc}-\ref{fig:proton}.

\begin{table*}
\caption {Posterior mean and $68\%$ credible uncertainties of the model
parameters}
\begin{tabular}{cccccccc}
\hline
\hline
                     &        Unit          &          PD          &          DC          &         DC2          &          DR          &         DR2          &         DRC         \\
\hline
       $D_0$         & $(10^{28}\mathrm{cm^2s^{-1}})$ &   $5.29 \pm 0.51$    &    $4.20 \pm 0.30$     &   $4.95 \pm 0.35$    &   $7.24 \pm 0.97$    &   $4.16 \pm 0.57$    &   $6.14 \pm 0.45$   \\
      $\delta$       &                      &  $0.471 \pm 0.006$   &  $0.588 \pm 0.013$   &  $0.591 \pm 0.011$   &   $0.380 \pm 0.007$   &   $0.500 \pm 0.012$    &  $0.478 \pm 0.013$  \\
       $z_h$         &   $(\mathrm{kpc})$   &   $6.61 \pm 0.98$    &    $10.90 \pm 1.60$    &    $10.80 \pm 1.30$    &   $5.93 \pm 1.13$    &   $5.02 \pm 0.86$    &    $12.70 \pm 1.40$   \\
       $v_A$         & $(\mathrm{km\,s^{-1}})$ &         ---          &         ---          &         ---          &   $38.5 \pm 1.3$    &   $18.4 \pm 2.0$   &   $43.2 \pm 1.2$  \\
      $dV_c/dz$        & $(\mathrm{km\,s^{-1}\,kpc^{-1}})$ &         ---          &   $5.36 \pm 0.64$   &  $5.02 \pm 0.55$   &         ---          &         ---          &   $11.99 \pm 1.26$  \\
       $R_0$         &   $(\mathrm{GV})$    &         ---          &         ---          &   $5.29 \pm 0.23$    &         ---          &         ---          &         ---         \\
       $\eta$        &                      &         ---          &         ---          &         ---          &         ---          &   $-1.28 \pm 0.22$   &         ---         \\
$\mathrm{log}(A_p)$\footnote{Propagated flux normalization at 100 GeV in unit of $\mathrm{cm^{-2}s^{-1}sr^{-1}MeV^{-1}}$} &                      &  $-8.334 \pm 0.003$  &  $-8.334 \pm 0.003$  &  $-8.336 \pm 0.003$  &  $-8.347 \pm 0.002$  &  $-8.334 \pm 0.002$  &  $-8.345 \pm 0.002$ \\
      $\nu_1$        &                      &   $2.44 \pm 0.01$    &   $2.45 \pm 0.01$    &   $2.43 \pm 0.01$    &   $1.69 \pm 0.02$    &   $2.04 \pm 0.03$    &   $1.82 \pm 0.02$   \\
      $\nu_2$        &                      &   $2.34 \pm 0.03$    &    $2.30 \pm 0.01$    &    $2.30 \pm 0.01$    &   $2.37 \pm 0.01$    &   $2.33 \pm 0.01$    &   $2.37 \pm 0.01$   \\
$\mathrm{log}(R_{br})$\footnote{Break rigidity of proton injection spectrum in unit of $\mathrm{MV}$} &                      &   $5.06 \pm 0.13$    &   $4.82 \pm 0.05$    &   $4.78 \pm 0.06$    &   $4.11 \pm 0.02$    &   $4.03 \pm 0.03$    &   $4.22 \pm 0.03$   \\
      $\Phi_0$       &   $(\mathrm{GV})$    &  $0.595 \pm 0.005$   &  $0.537 \pm 0.006$   &  $0.419 \pm 0.005$   &   $0.180 \pm 0.008$   &   $0.290 \pm 0.014$   &   $0.220 \pm 0.008$  \\
      $\Phi_1$       &   $(\mathrm{GV})$    &  $0.495 \pm 0.011$   &  $0.485 \pm 0.011$   &  $0.472 \pm 0.012$   &  $0.487 \pm 0.011$   &  $0.485 \pm 0.011$   &  $0.482 \pm 0.013$  \\
$\chi^2/\mathrm{dof}$ &                      &      748.6/463       &       591.0/462        &      494.6/461       &      438.8/462       &       341.0/461        &      380.5/461      \\\hline
\end{tabular}
\label{table:para}
\end{table*}

The fittings show that the models with reacceleration (DR, DR2, and DRC) 
can fit the B/C and proton data well, while the other three 
non-reacceleration models fit the data relatively poorly\footnote{Note,
however, the study of CR electrons and positrons results in a different
conclusion, i.e., the convection models are more favored than the
reacceleration models \cite{2015PhRvD..91f3508L}.}. The reduced
chi-squared values are all smaller than 1 for the three reacceleration 
models. For the non-reacceleration models, the $\chi^2$ values indicate
$p-$values of $\sim7.8\times10^{-16}$, $4.3\times10^{-5}$, and $0.14$
for the PD, DC, and DC2 models, respectively. From Fig. \ref{fig:bc}
we can see that the predicted B/C ratios for non-reacceleration models
do not match the low energy ($E_k\lesssim1$ GeV/n) data well. This
is perhaps due to larger solar modulation potentials for non-reacceleration 
models, which are required by the proton data. These results illustrate 
the importance of including the low energy ACE data of B/C and the primary
CR flux data when studying the propagation of CRs.

There is a clear degeneracy between $D_0$ and $z_h$. This is because the
B/C data can only constrain $D_0/z_h$ effectively \cite{2001ApJ...555..585M,
2015JCAP...09..049J}. The unstable-to-stable secondary ratio is expected
to break such a degeneracy. However, the current $^{10}$Be/$^9$Be ratio
data are of relatively poor quality. The $95\%$ credible region of $D_0$ 
is $[5.2,9.2]\times10^{28}$ cm$^2$ s$^{-1}$ for the DR model, and the 
cresponding value of $z_h$ is $[3.7,8.2]$ kpc. As a comparison, they are 
$[5.45,11.20]\times10^{28}$ cm$^2$ s$^{-1}$, and $[3.2,8.6]$ kpc in Ref. 
\cite{2011ApJ...729..106T}. Our results improve moderately compared with 
that of Ref. \cite{2011ApJ...729..106T}. Through analyzing the
synchrotron radiation and the electron/positron fluxes, Di Bernardo et al.
also found a relatively large propagation halo height ($z_h>2$ kpc;
\cite{2013JCAP...03..036D}), which is consistent with our results.

There are some other correlations among the propagation parameters.
For example, for the DC and DC2 scenario, an anti-correlation between
$D_0$ and $dV_c/dz$ can be found (Figs. \ref{fig:tri_DC} and 
\ref{fig:tri_DC2}). This can be understood that, a larger convection
velocity tends to blow the particles away from the disk, resulting in
a lower flux, which can be compensated by a longer propagation time
(hence a smaller $D_0$). A positive correlation between $\delta$ and
$dV_c/dz$ can be understood similarly. Since the convection is 
only important for low energy particles, a larger convection velocity
will lead to harder spectra of the CR fluxes and B/C ratio, which
can be compensated by a larger value of $\delta$. For the DR2 scenario, 
we find anti-correlations between $v_A$ and $\delta$, $\eta$ and
$\delta$, and positive correlation between $v_A$ and $\eta$. A larger
$v_A$ value gives softer spectra of the CR fluxes and B/C ratio, and
hence suggesting a smaller $\delta$. The anti-correlation between
$\eta$ and $\delta$ can be understood as: a smaller $\eta$ (note that
$\eta<0$) gives a larger diffusion coefficient at low energies, and
results in harder spectra after the propagation. A larger value of
$\delta$ is then able to compensate such an effect. 

The slope $\delta$ of the diffusion coefficient is well constrained 
(with statistical uncertainty of a few percents) given the model setting. 
However, there are relative large differences among different model
configurations. For the reacceleration models, $\delta$ is about $0.38$
for the DR model, and about $0.5$ for the DR2/DRC models. For the DC/DC2
models, $\delta$ is even larger (about $0.6$). These results can be 
understood via the correlations between $\delta$ and other parameters
as described above. The fitting to the B/C ratio above 65 GV gives a 
slope of $-0.333$ \cite{2016PhRvL.117w1102A}.
Our results show that in specific models the value of $\delta$ may
differ from that directly inferred from the data. This is because,
on one hand, the low energy spectrum of the B/C ratio depends on 
propagation models, and on the other hand, the uncertainties of high 
energy data are relatively large. It is currently difficult to
distinguish the Kolmogrov ($\delta=1/3$; \cite{1941DoSSR..30..301K}) 
and the Kraichnan ($\delta=1/2$; \cite{1965PhFl....8.1385K}) type of 
interstellar turbulence. Nevertheless, we find that for some of the
propagation model settings, such as the DR2 and DRC models, the Kraichnan
type of turbulence is favored. For the DR model, the fitting value of
$\delta$ is closer to, but still different from, that predicted by the 
Kolmogrov theory.

For reacceleration models, the Alfven velocity $v_A$ is about 38 km 
s$^{-1}$ for the DR model, which decreases (increases) to about 18 (43) 
km s$^{-1}$ for the DR2 (DRC) model. The major effect of reacceleration
is to produce a ``GeV bump'' of the CR flux and B/C ratio. For the DR2
model, a larger $\delta$ gives higher B/C ratio at lower energies,
and hence a smaller reacceleration effect is needed. This can also be
seen from the anti-correlation between $v_A$ and $\delta$ (Fig.
\ref{fig:tri_DR2}). The effect of convection is, however, opposite from
that of reacceleration. Therefore for the DRC model, a larger value of
$v_A$ is favored given a non-zero value of $dV_c/dz$.

A break of the injection spectrum around $10-20$ GV is favored in the 
reacceleration models. Such a break is required to fit the proton fluxes,
in order to reduce the ``GeV bump'' produced by the reacceleration. 
Such a break is not necessary for the non-reacceleration models.
Nevertheless, we find that a spectral hardening with a change of the
slope of $\sim0.10-0.15$ is favored by the fitting. Such a break enables
a better fit to the high energy proton flux by the AMS-02 which shows
a spectral hardening above $\sim330$ GV. The break rigidity is not
exactly the same as that obtained directly from the data, because the
low energy spectral behavior also enters in the fitting. 

As for the solar modulation, we find that the time-dependent term of the 
modulation potential, $\Phi_1$, is similar for all models. It reflects the 
differences of the proton fluxes at different time. The platform term 
$\Phi_0$ differ from each other. In general, non-reacceleration models need 
remarkably larger $\Phi_0$ to accommodate the low energy data of protons.

\subsection{Positrons}

\begin{figure*}[!htb]
\includegraphics[width=\textwidth]{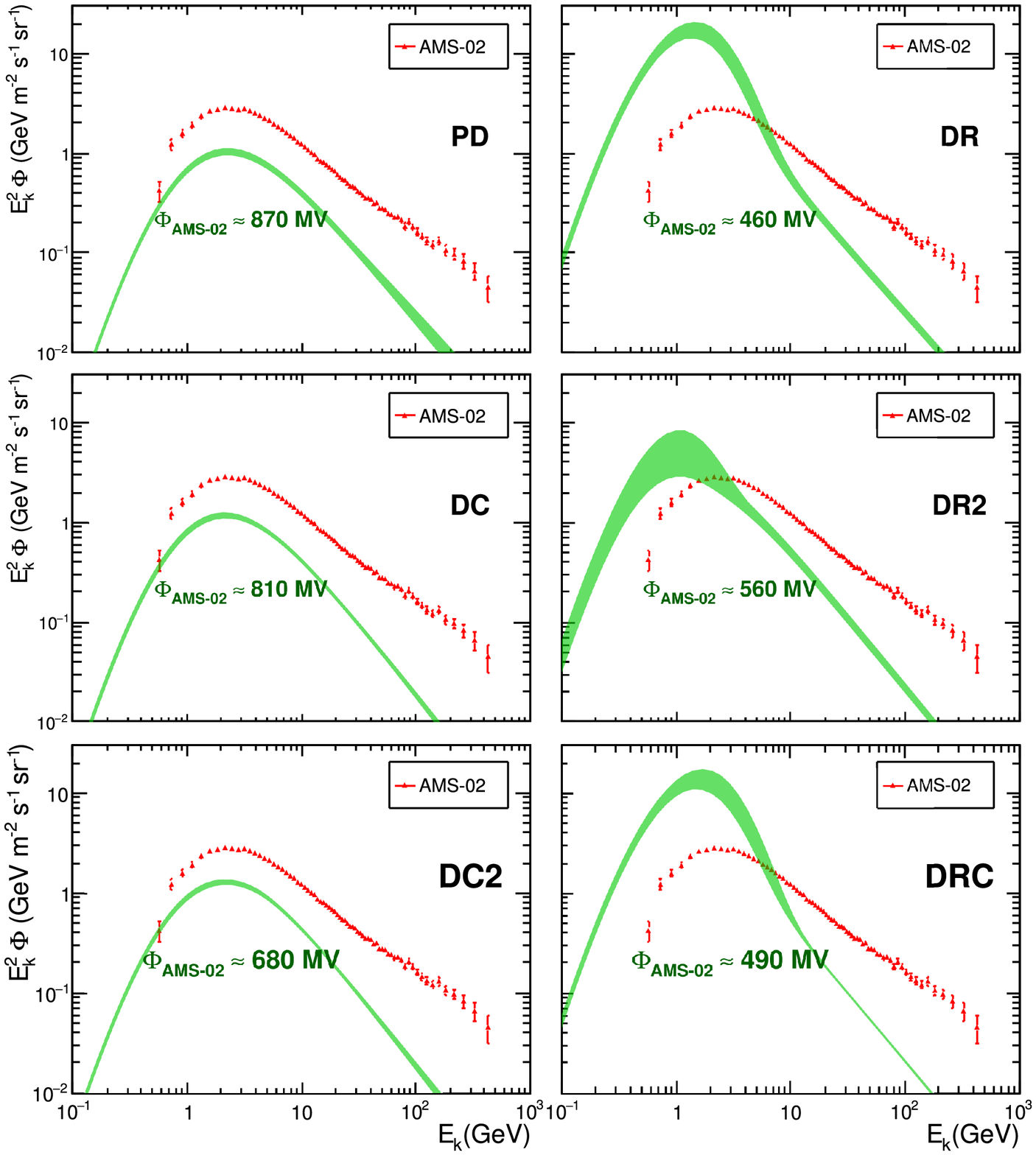}
\caption{Predicted $2\sigma$ bands of the positron spectra, compared with
the AMS-02 measurements \cite{2014PhRvL.113l1102A}.
\label{fig:positron}}
\end{figure*}

The fluxes of secondary positrons can be calculated self-consistently given 
the fitting propagation and source parameters. Fig. \ref{fig:positron} 
shows the expected $2\sigma$ bands of positron fluxes, compared with the 
AMS-02 data \cite{2014PhRvL.113l1102A}. We find that the reacceleration 
models which fit the B/C and proton data well would result in a remarkable 
bump at $\sim$GeV energies and exceed the data significantly. This is 
consistent with that found in earlier studies \cite{2002ApJ...565..280M,
2011ApJ...729..106T}. For the non-reacceleration models, on the other 
hand, the expected positron fluxes are lower than the data by a factor of 
$\sim2-3$. These results indicate that the production and propagation of 
positrons may be significantly different from that of the CR nuclei.

For all these models, the predicted positron spectra at high energies 
($\gtrsim10$ GeV) are much softer than that of the data, which indicate
the existence of primary positron sources, e.g., pulsars
\cite{1970ApJ...162L.181S,1987ICRC....2...92H,2001A&A...368.1063Z}.

\subsection{Antiprotons}

\begin{figure*}[!htb]
\includegraphics[width=\textwidth]{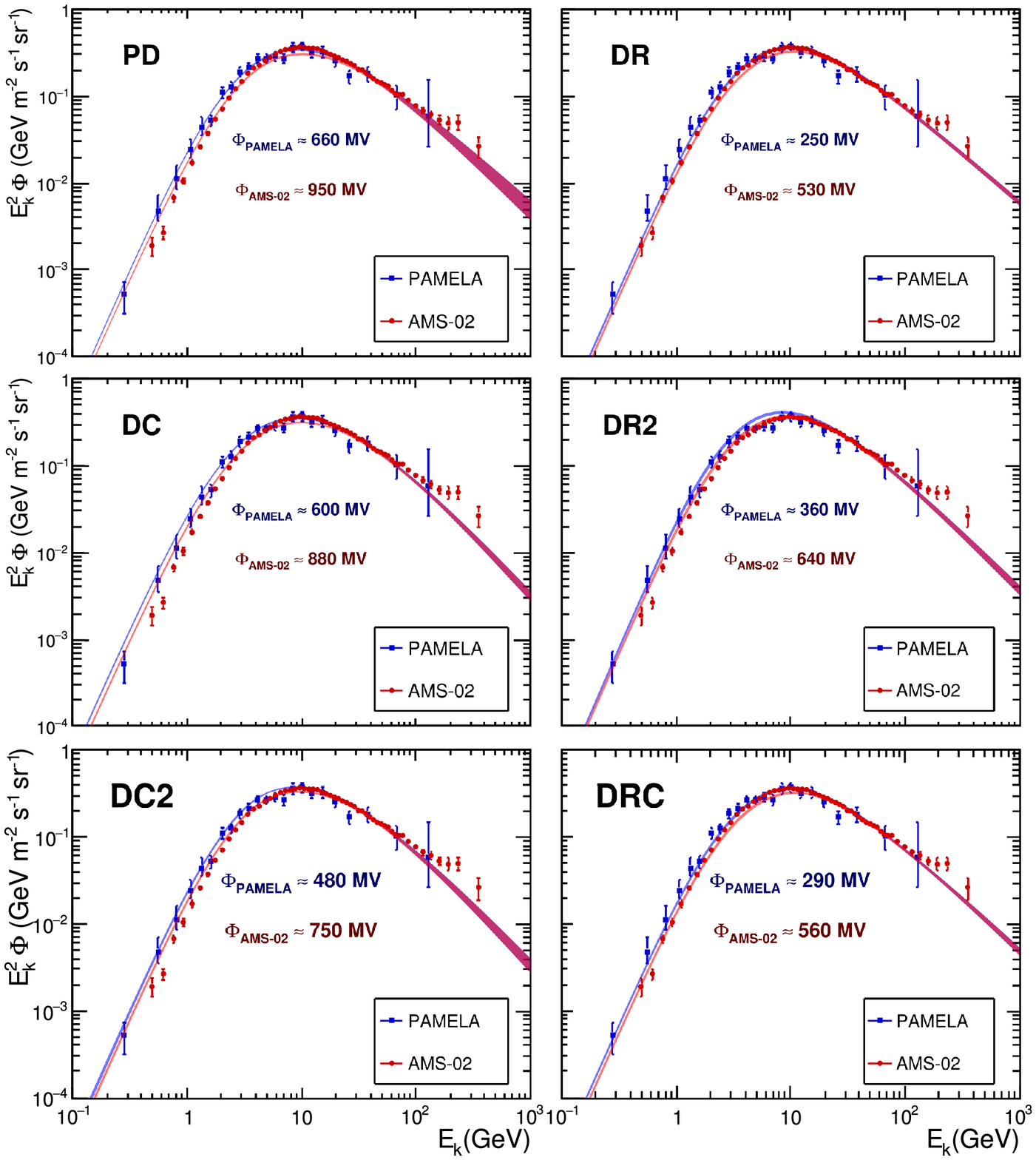}
\caption{Predicted $2\sigma$ bands of the antiproton spectra, compared
with the PAMELA \cite{2010PhRvL.105l1101A} and AMS-02 data
\cite{2016PhRvL.117i1103A}.
\label{fig:antiproton}}
\end{figure*}

Fig. \ref{fig:antiproton} shows the results of antiprotons from the
models, compared with the PAMELA \cite{2010PhRvL.105l1101A} and AMS-02
\cite{2016PhRvL.117i1103A} measurements. We find that the model predictions 
are roughly consistent with the data. More detailed comparison shows that 
in general the non-reacceleration model predictions match the data better 
than the reacceleration models. For the DR and DRC models, there are slight 
deficits of low energy ($\lesssim10$ GeV) antiprotons compared with the 
data. The DR2 model can marginally fit the data. The prediction of the 
DC2 model is consistent with the data. For the PD and DC models, however, 
they slightly under-predict antiprotons around 10 GeV and over-predict 
lower energy antiprotons. At the high energy end ($E\gtrsim100$ GeV), 
there might be excesses of the data (see also \cite{2016arXiv161101983H,
2016arXiv161209501L,2017arXiv170102263F,2017arXiv170104406C}). 
For models with larger $\delta$ values such as the DC, DC2, and DR2 models, 
the excesses are remarkable. For the other three models with relatively 
smaller $\delta$ values such excesses are less significant.

\section{Discussion}

\subsection{The discrepancy between non-reacceleration models and
CR nuclei data}

It seems that the non-reacceleration models have difficulty to fit
the proton fluxes and the B/C ratio simultaneously. We find that for
the non-reacceleration models the required solar modulation potential
($\Phi_0$) is significantly higher than that of reacceleration models,
which results in poor fittings to the low energy B/C data of ACE.
To test that whether such a discrepancy is due to the difference of 
solar modulation between protons and heavy nuclei, we do similar 
fittings using the preliminary Carbon flux by AMS-02 
\cite{2016-AMS02-CERN} instead of the proton fluxes. We find similar 
conclusion as above, which means that the difference of solar modulation 
between protons and heavy nuclei is not the major reason of this 
discrepancy.

Another possible reason is the injection spectrum of CRs. For the three 
non-reacceleration models the injection spectrum at low rigidities is 
proportional to $R^{-(2.4-2.5)}$, which is quite soft compared with 
that of the reacceleration models $R^{-(1.7-2.0)}$. Even though we
enable a break of the low energy spectrum of all the models, the
fitting results turn out to favor a high energy hardening instead
of a low energy break for the non-reacceleration models. We have added 
another break in the injection spectrum of Eq. (\ref{injection_equation}), 
and redo the fittings. Still no effective improvement of the fittings 
is found.

\subsection{The discrepancy between all models and the positron data}

Our results show that the reacceleration models would over-predict low
energy ($\sim$GeV) positrons compared with the measurements, while the 
non-reacceleration models tend to under-predict positrons. Similar
results for reacceleration models have also been obtained in Ref.
\cite{2013JCAP...03..036D}. One kind of uncertainty is the hadronic 
$pp$-interaction. In this work we use the parameterization of positron 
production in $pp$-interaction of Ref. \cite{2006ApJ...647..692K}. 
As illustrated in Ref. \cite{2009A&A...501..821D}, some other 
parameterizations would give a positron yield spectrum differing 
by a factor of $\lesssim2$ in a certain energy range. However, the 
uncertainty of the hadronic interaction may not be able to fully solve 
this discrepancy, especially for the reacceleration models. The other 
models adopted in Ref. \cite{2009A&A...501..821D} give even more positrons 
between GeV and TeV, which makes the reacceleration models exceed the data 
even more. Therefore our results indicate that the propagation of CR nuclei
and leptons, either in the Galaxy or in the heliosphere, might be
different. Given the very efficient energy losses of leptons, they may 
experience large fluctuations in the Galaxy \cite{1998ApJ...507..327P}.
The solar modulation effects may also be different between nuclei and
leptons due to their distinct mass-to-charge ratios. The charge-sign
dependent solar modulation may take effect either
\cite{1996ApJ...464..507C,2012AdSpR..49.1587D,2013PhRvL.110h1101M,
2014SoPh..289..391P,2016CoPhC.207..386K}.

\subsection{The Voyager-1 measurements in outer heliosphere}

The Voyager-1 spacecraft has traveled by more than 100 astronomical 
units from the Earth. It has been thought to approach the edge of the 
heliosphere since a sudden drop of the intensity of low energy ions and 
an abrupt increase of the CR intensity from outside the heliosphere were
observed \cite{2013Sci...341..150S}. The measured CR flux by Voyager-1 
can thus be believed to be a direct measurement of the local interstellar 
CRs. The Voyager-1 data would be helpful in better constraining the source 
injection parameters as well as the solar modulation parameters. However, 
as shown in Ref. \cite{2016ApJ...831...18C}, the very low energy 
($\lesssim50$ MeV/n) B/C spectrum measured by Voyager-1 is difficult to 
be modelled in various models. Further tuning of the modelling and/or 
better understanding about the measurements may be necessary. The Voyager-1 
data will be included in future studies.

%\begin{figure}[!htb]
%\includegraphics[width=0.45\textwidth]{proton_together.eps}
%\caption{The LIS proton spectra from all the models in this work compared 
%with the data of Voyager 1 \cite{2016ApJ...831...18C} and AMS-02 
%\cite{2015PhRvL.114q1103A}.}
%\label{fig:voyager}
%\end{figure}

\subsection{Reacceleration models and antiprotons}

The reacceleration models would generally under-estimate the low energy
antiproton fluxes. Several kinds of scenarios were proposed to explain
this. In Ref. \cite{2003ApJ...586.1050M} it was proposed that a local
and fresh source, probably associated with the {\it Local Bubble}, might
produce additional low energy primaries and hence decrease the measured
secondary-to-primary nuclei ratio. The annihilation of several tens of
GeV dark matter particles may also be responsible for the low energy
excess of antiprotons \cite{2015JCAP...03..021H,2016arXiv161003840C,
2016arXiv161003071C}. Alternatively, an empirical adjustement of the
velocity-dependence of the diffusion coefficient with a $\beta^{\eta}$
term, i.e., the DR2 model in this work, was suggested to be able to
explain the B/C and antiproton data \cite{2010APh....34..274D}. In this
treatment a larger $\delta$ value and a weaker reacceleration effect is
required, which enables more production of low energy secondary particles
(both Boron and antiprotons). As shown in Fig. \ref{fig:antiproton}, the 
DR2 model does improve the fitting. However, the physical motivation for 
such a term is not well justified. Finally, the uncertainties of the 
production cross section of antiprotons make this problem still inconclusive 
\cite{2001ApJ...563..172D,2015JCAP...09..023G,2016arXiv161204001L}.

\section{Conclusion}

In this work we adopt the precise measurements of the B/C ratio and the
time-dependent proton fluxes by AMS-02 and PAMELA to constrain the
injection and propagation parameters of Galactic CRs. We employ a
self-consistent treatment of the solar modulation by means of a linear
correlation of the modulation potentials with solar activities. We have
carried out a comprehensive study of a series of CR propagation models,
including the PD, DR, DC, DRC, and two variants of the DR and DC models.
The predictions of secondary positrons and antiprotons based on the
fitting parameters are calculated and compared with the data. 

We summarize the comparison of various models with different data sets 
in Table \ref{table:summary}. It is shown that no model can match all 
these data simultaneously, which suggests that the actual case for the 
origin, propagation, and interaction of CRs is more complicated than our
current understanding. For the CR nuclei only, we find that the DR2 model
may give the best match to all the data. However, the phenomenological
modification of the diffusion coefficient (the $\beta^{\eta}$ term) may 
need to be understood further \cite{2010APh....34..274D}.

\begin{table}[!htb]
\caption {Summary of different propagation models versus the data}
\begin{tabular}{lcccc}
\hline \hline
 & B/C \& protons &  positrons & antiprotons \\
\hline
PD & Poor & Too few & Fair \\
DC & Poor & Too few & Fair \\
DC2 & Poor & Too few & Good \\
DR & Good & Too many & Slightly few \\
DR2 & Good & Too many & Fair \\
DRC & Good & Too many & Slightly few \\
\hline
\hline
\end{tabular}
\label{table:summary}
\end{table}

We list our main conclusion as follows.
\begin{itemize}

\item The reacceleration models (DR, DR2, and DRC) can fit both the B/C
and proton fluxes well, while non-reacceleration models (PD, DC, and DC2)
can not. The failure of non-reacceleration models can not be simply
ascribed to the differences of solar modulation or the source injection
spectra between protons and heavier nuclei.

\item The statistical uncertainties of the propagation parameters are
constrained to a level of $10\%-20\%$, thanks to the precise measurements 
of CR data by AMS-02. However, there are relatively large differences
(up to a factor of $\sim2$) among different model settings. 

\item For reacceleration models, the value of $\delta$ is found to be 
about $0.38-0.50$, which slightly favor the Kraichnan type of interstellar 
turbulence. 

\item The reacceleration models will over-produce positrons but 
under-produce (except DR2) antiprotons in general. The non-reacceleration 
models, on the other hand, predict fewer positrons and (marginally) 
consistent antiprotons when compared with the measurements.

\item Our results suggest that there are significant differences of the
propagation in either the Milky Way or the heliosphere between nuclei
and leptons.

\end{itemize}

With more and more precise data available, we are able to investigate
the CR-related problems in great detail. It turns out that the problem
seems to be more complicated than what we expected based on the rough
measurements in the past. The final understanding of the propagation
of CRs may need not only the CR data themselves but also the full
improvements of the understanding of the astrophysical ingradients
of the Milky Way, as well as the nuclear and hadronic interactions.

\acknowledgments
We thank the ACE CRIS instrument team and the ACE Science Center for
providing the ACE data. This work is supported by the National Key 
Research and Development Program of China (No. 2016YFA0400200), the 
National Natural Science Foundation of China (No. 11475191), and the 
100 Talents program of Chinese Academy of Sciences.

\bibliographystyle{apsrev}
\bibliography{/home/yuanq/work/cygnus/tex/refs}

\begin{thebibliography}{90}
\expandafter\ifx\csname natexlab\endcsname\relax\def\natexlab#1{#1}\fi
\expandafter\ifx\csname bibnamefont\endcsname\relax
  \def\bibnamefont#1{#1}\fi
\expandafter\ifx\csname bibfnamefont\endcsname\relax
  \def\bibfnamefont#1{#1}\fi
\expandafter\ifx\csname citenamefont\endcsname\relax
  \def\citenamefont#1{#1}\fi
\expandafter\ifx\csname url\endcsname\relax
  \def\url#1{\texttt{#1}}\fi
\expandafter\ifx\csname urlprefix\endcsname\relax\def\urlprefix{URL }\fi
\providecommand{\bibinfo}[2]{#2}
\providecommand{\eprint}[2][]{\url{#2}}

\bibitem[{\citenamefont{{Gaisser}}(1990)}]{1990cup..book.....G}
\bibinfo{author}{\bibfnamefont{T.~K.} \bibnamefont{{Gaisser}}},
  \emph{\bibinfo{title}{{Cosmic rays and particle physics}}}
  (\bibinfo{publisher}{Cambridge and New York, Cambridge University Press,
  1990, 292 p.}, \bibinfo{year}{1990}).

\bibitem[{\citenamefont{{Strong} et~al.}(2007)\citenamefont{{Strong},
  {Moskalenko}, and {Ptuskin}}}]{2007ARNPS..57..285S}
\bibinfo{author}{\bibfnamefont{A.~W.} \bibnamefont{{Strong}}},
  \bibinfo{author}{\bibfnamefont{I.~V.} \bibnamefont{{Moskalenko}}},
  \bibnamefont{and} \bibinfo{author}{\bibfnamefont{V.~S.}
  \bibnamefont{{Ptuskin}}}, \bibinfo{journal}{Annual Review of Nuclear and
  Particle Science} \textbf{\bibinfo{volume}{57}}, \bibinfo{pages}{285}
  (\bibinfo{year}{2007}), \eprint{astro-ph/0701517}.

\bibitem[{\citenamefont{{Berezinskii} et~al.}(1990)\citenamefont{{Berezinskii},
  {Bulanov}, {Dogiel}, and {Ptuskin}}}]{1990acr..book.....B}
\bibinfo{author}{\bibfnamefont{V.~S.} \bibnamefont{{Berezinskii}}},
  \bibinfo{author}{\bibfnamefont{S.~V.} \bibnamefont{{Bulanov}}},
  \bibinfo{author}{\bibfnamefont{V.~A.} \bibnamefont{{Dogiel}}},
  \bibnamefont{and} \bibinfo{author}{\bibfnamefont{V.~S.}
  \bibnamefont{{Ptuskin}}}, \emph{\bibinfo{title}{{Astrophysics of cosmic
  rays}}} (\bibinfo{publisher}{Amsterdam: North-Holland, 1990, edited by
  Ginzburg, V.L.}, \bibinfo{year}{1990}).

\bibitem[{\citenamefont{{Webber} et~al.}(1992)\citenamefont{{Webber}, {Lee},
  and {Gupta}}}]{1992ApJ...390...96W}
\bibinfo{author}{\bibfnamefont{W.~R.} \bibnamefont{{Webber}}},
  \bibinfo{author}{\bibfnamefont{M.~A.} \bibnamefont{{Lee}}}, \bibnamefont{and}
  \bibinfo{author}{\bibfnamefont{M.}~\bibnamefont{{Gupta}}},
  \bibinfo{journal}{\apj} \textbf{\bibinfo{volume}{390}}, \bibinfo{pages}{96}
  (\bibinfo{year}{1992}).

\bibitem[{\citenamefont{{Bloemen} et~al.}(1993)\citenamefont{{Bloemen},
  {Dogel'}, {Dorman}, and {Ptuskin}}}]{1993A&A...267..372B}
\bibinfo{author}{\bibfnamefont{J.~B.~G.~M.} \bibnamefont{{Bloemen}}},
  \bibinfo{author}{\bibfnamefont{V.~A.} \bibnamefont{{Dogel'}}},
  \bibinfo{author}{\bibfnamefont{V.~L.} \bibnamefont{{Dorman}}},
  \bibnamefont{and} \bibinfo{author}{\bibfnamefont{V.~S.}
  \bibnamefont{{Ptuskin}}}, \bibinfo{journal}{Astronomy and Astrophys.}
  \textbf{\bibinfo{volume}{267}}, \bibinfo{pages}{372} (\bibinfo{year}{1993}).

\bibitem[{\citenamefont{{Maurin} et~al.}(2001)\citenamefont{{Maurin}, {Donato},
  {Taillet}, and {Salati}}}]{2001ApJ...555..585M}
\bibinfo{author}{\bibfnamefont{D.}~\bibnamefont{{Maurin}}},
  \bibinfo{author}{\bibfnamefont{F.}~\bibnamefont{{Donato}}},
  \bibinfo{author}{\bibfnamefont{R.}~\bibnamefont{{Taillet}}},
  \bibnamefont{and} \bibinfo{author}{\bibfnamefont{P.}~\bibnamefont{{Salati}}},
  \bibinfo{journal}{\apj} \textbf{\bibinfo{volume}{555}}, \bibinfo{pages}{585}
  (\bibinfo{year}{2001}).

\bibitem[{\citenamefont{{Maurin} et~al.}(2002)\citenamefont{{Maurin},
  {Taillet}, and {Donato}}}]{2002A&A...394.1039M}
\bibinfo{author}{\bibfnamefont{D.}~\bibnamefont{{Maurin}}},
  \bibinfo{author}{\bibfnamefont{R.}~\bibnamefont{{Taillet}}},
  \bibnamefont{and} \bibinfo{author}{\bibfnamefont{F.}~\bibnamefont{{Donato}}},
  \bibinfo{journal}{Astronomy and Astrophys.} \textbf{\bibinfo{volume}{394}},
  \bibinfo{pages}{1039} (\bibinfo{year}{2002}).

\bibitem[{\citenamefont{{Shibata} et~al.}(2004)\citenamefont{{Shibata},
  {Hareyama}, {Nakazawa}, and {Saito}}}]{2004ApJ...612..238S}
\bibinfo{author}{\bibfnamefont{T.}~\bibnamefont{{Shibata}}},
  \bibinfo{author}{\bibfnamefont{M.}~\bibnamefont{{Hareyama}}},
  \bibinfo{author}{\bibfnamefont{M.}~\bibnamefont{{Nakazawa}}},
  \bibnamefont{and} \bibinfo{author}{\bibfnamefont{C.}~\bibnamefont{{Saito}}},
  \bibinfo{journal}{\apj} \textbf{\bibinfo{volume}{612}}, \bibinfo{pages}{238}
  (\bibinfo{year}{2004}).

\bibitem[{\citenamefont{{Strong} and {Moskalenko}}(1998)}]{1998ApJ...509..212S}
\bibinfo{author}{\bibfnamefont{A.~W.} \bibnamefont{{Strong}}} \bibnamefont{and}
  \bibinfo{author}{\bibfnamefont{I.~V.} \bibnamefont{{Moskalenko}}},
  \bibinfo{journal}{\apj} \textbf{\bibinfo{volume}{509}}, \bibinfo{pages}{212}
  (\bibinfo{year}{1998}), \eprint{astro-ph/9807150}.

\bibitem[{\citenamefont{{Moskalenko} and {Strong}}(1998)}]{1998ApJ...493..694M}
\bibinfo{author}{\bibfnamefont{I.~V.} \bibnamefont{{Moskalenko}}}
  \bibnamefont{and} \bibinfo{author}{\bibfnamefont{A.~W.}
  \bibnamefont{{Strong}}}, \bibinfo{journal}{\apj}
  \textbf{\bibinfo{volume}{493}}, \bibinfo{pages}{694} (\bibinfo{year}{1998}),
  \eprint{astro-ph/9710124}.

\bibitem[{\citenamefont{{Evoli} et~al.}(2008)\citenamefont{{Evoli}, {Gaggero},
  {Grasso}, and {Maccione}}}]{2008JCAP...10..018E}
\bibinfo{author}{\bibfnamefont{C.}~\bibnamefont{{Evoli}}},
  \bibinfo{author}{\bibfnamefont{D.}~\bibnamefont{{Gaggero}}},
  \bibinfo{author}{\bibfnamefont{D.}~\bibnamefont{{Grasso}}}, \bibnamefont{and}
  \bibinfo{author}{\bibfnamefont{L.}~\bibnamefont{{Maccione}}},
  \bibinfo{journal}{\jcap} \textbf{\bibinfo{volume}{10}}, \bibinfo{pages}{18}
  (\bibinfo{year}{2008}), \eprint{0807.4730}.

\bibitem[{\citenamefont{{Swordy} et~al.}(1990)\citenamefont{{Swordy},
  {Mueller}, {Meyer}, {L'Heureux}, and {Grunsfeld}}}]{1990ApJ...349..625S}
\bibinfo{author}{\bibfnamefont{S.~P.} \bibnamefont{{Swordy}}},
  \bibinfo{author}{\bibfnamefont{D.}~\bibnamefont{{Mueller}}},
  \bibinfo{author}{\bibfnamefont{P.}~\bibnamefont{{Meyer}}},
  \bibinfo{author}{\bibfnamefont{J.}~\bibnamefont{{L'Heureux}}},
  \bibnamefont{and} \bibinfo{author}{\bibfnamefont{J.~M.}
  \bibnamefont{{Grunsfeld}}}, \bibinfo{journal}{\apj}
  \textbf{\bibinfo{volume}{349}}, \bibinfo{pages}{625} (\bibinfo{year}{1990}).

\bibitem[{\citenamefont{{Mueller} et~al.}(1991)\citenamefont{{Mueller},
  {Swordy}, {Meyer}, {L'Heureux}, and {Grunsfeld}}}]{1991ApJ...374..356M}
\bibinfo{author}{\bibfnamefont{D.}~\bibnamefont{{Mueller}}},
  \bibinfo{author}{\bibfnamefont{S.~P.} \bibnamefont{{Swordy}}},
  \bibinfo{author}{\bibfnamefont{P.}~\bibnamefont{{Meyer}}},
  \bibinfo{author}{\bibfnamefont{J.}~\bibnamefont{{L'Heureux}}},
  \bibnamefont{and} \bibinfo{author}{\bibfnamefont{J.~M.}
  \bibnamefont{{Grunsfeld}}}, \bibinfo{journal}{\apj}
  \textbf{\bibinfo{volume}{374}}, \bibinfo{pages}{356} (\bibinfo{year}{1991}).

\bibitem[{\citenamefont{{Yanasak} et~al.}(2001)\citenamefont{{Yanasak},
  {Wiedenbeck}, {Mewaldt}, {Davis}, {Cummings}, {George}, {Leske}, {Stone},
  {Christian}, {von Rosenvinge} et~al.}}]{2001ApJ...563..768Y}
\bibinfo{author}{\bibfnamefont{N.~E.} \bibnamefont{{Yanasak}}},
  \bibnamefont{et~al.}, \bibinfo{journal}{\apj} \textbf{\bibinfo{volume}{563}},
  \bibinfo{pages}{768} (\bibinfo{year}{2001}).

\bibitem[{\citenamefont{{Lionetto} et~al.}(2005)\citenamefont{{Lionetto},
  {Morselli}, and {Zdravkovic}}}]{2005JCAP...09..010L}
\bibinfo{author}{\bibfnamefont{A.~M.} \bibnamefont{{Lionetto}}},
  \bibinfo{author}{\bibfnamefont{A.}~\bibnamefont{{Morselli}}},
  \bibnamefont{and}
  \bibinfo{author}{\bibfnamefont{V.}~\bibnamefont{{Zdravkovic}}},
  \bibinfo{journal}{\jcap} \textbf{\bibinfo{volume}{9}}, \bibinfo{pages}{10}
  (\bibinfo{year}{2005}), \eprint{astro-ph/0502406}.

\bibitem[{\citenamefont{{Ave} et~al.}(2009)\citenamefont{{Ave}, {Boyle},
  {H{\"o}ppner}, {Marshall}, and {M{\"u}ller}}}]{2009ApJ...697..106A}
\bibinfo{author}{\bibfnamefont{M.}~\bibnamefont{{Ave}}},
  \bibinfo{author}{\bibfnamefont{P.~J.} \bibnamefont{{Boyle}}},
  \bibinfo{author}{\bibfnamefont{C.}~\bibnamefont{{H{\"o}ppner}}},
  \bibinfo{author}{\bibfnamefont{J.}~\bibnamefont{{Marshall}}},
  \bibnamefont{and}
  \bibinfo{author}{\bibfnamefont{D.}~\bibnamefont{{M{\"u}ller}}},
  \bibinfo{journal}{\apj} \textbf{\bibinfo{volume}{697}}, \bibinfo{pages}{106}
  (\bibinfo{year}{2009}), \eprint{0810.2972}.

\bibitem[{\citenamefont{{Pato} et~al.}(2010)\citenamefont{{Pato}, {Hooper}, and
  {Simet}}}]{2010JCAP...06..022P}
\bibinfo{author}{\bibfnamefont{M.}~\bibnamefont{{Pato}}},
  \bibinfo{author}{\bibfnamefont{D.}~\bibnamefont{{Hooper}}}, \bibnamefont{and}
  \bibinfo{author}{\bibfnamefont{M.}~\bibnamefont{{Simet}}},
  \bibinfo{journal}{\jcap} \textbf{\bibinfo{volume}{6}}, \bibinfo{pages}{22}
  (\bibinfo{year}{2010}), \eprint{1002.3341}.

\bibitem[{\citenamefont{{di Bernardo} et~al.}(2010)\citenamefont{{di Bernardo},
  {Evoli}, {Gaggero}, {Grasso}, and {Maccione}}}]{2010APh....34..274D}
\bibinfo{author}{\bibfnamefont{G.}~\bibnamefont{{di Bernardo}}},
  \bibinfo{author}{\bibfnamefont{C.}~\bibnamefont{{Evoli}}},
  \bibinfo{author}{\bibfnamefont{D.}~\bibnamefont{{Gaggero}}},
  \bibinfo{author}{\bibfnamefont{D.}~\bibnamefont{{Grasso}}}, \bibnamefont{and}
  \bibinfo{author}{\bibfnamefont{L.}~\bibnamefont{{Maccione}}},
  \bibinfo{journal}{Astroparticle Physics} \textbf{\bibinfo{volume}{34}},
  \bibinfo{pages}{274} (\bibinfo{year}{2010}), \eprint{0909.4548}.

\bibitem[{\citenamefont{{Obermeier} et~al.}(2012)\citenamefont{{Obermeier},
  {Boyle}, {H{\"o}randel}, and {M{\"u}ller}}}]{2012ApJ...752...69O}
\bibinfo{author}{\bibfnamefont{A.}~\bibnamefont{{Obermeier}}},
  \bibinfo{author}{\bibfnamefont{P.}~\bibnamefont{{Boyle}}},
  \bibinfo{author}{\bibfnamefont{J.}~\bibnamefont{{H{\"o}randel}}},
  \bibnamefont{and}
  \bibinfo{author}{\bibfnamefont{D.}~\bibnamefont{{M{\"u}ller}}},
  \bibinfo{journal}{\apj} \textbf{\bibinfo{volume}{752}}, \bibinfo{eid}{69}
  (\bibinfo{year}{2012}), \eprint{1204.6188}.

\bibitem[{\citenamefont{{Putze} et~al.}(2009)\citenamefont{{Putze}, {Derome},
  {Maurin}, {Perotto}, and {Taillet}}}]{2009A&A...497..991P}
\bibinfo{author}{\bibfnamefont{A.}~\bibnamefont{{Putze}}},
  \bibinfo{author}{\bibfnamefont{L.}~\bibnamefont{{Derome}}},
  \bibinfo{author}{\bibfnamefont{D.}~\bibnamefont{{Maurin}}},
  \bibinfo{author}{\bibfnamefont{L.}~\bibnamefont{{Perotto}}},
  \bibnamefont{and}
  \bibinfo{author}{\bibfnamefont{R.}~\bibnamefont{{Taillet}}},
  \bibinfo{journal}{\aap} \textbf{\bibinfo{volume}{497}}, \bibinfo{pages}{991}
  (\bibinfo{year}{2009}), \eprint{0808.2437}.

\bibitem[{\citenamefont{{Putze} et~al.}(2010)\citenamefont{{Putze}, {Derome},
  and {Maurin}}}]{2010A&A...516A..66P}
\bibinfo{author}{\bibfnamefont{A.}~\bibnamefont{{Putze}}},
  \bibinfo{author}{\bibfnamefont{L.}~\bibnamefont{{Derome}}}, \bibnamefont{and}
  \bibinfo{author}{\bibfnamefont{D.}~\bibnamefont{{Maurin}}},
  \bibinfo{journal}{\aap} \textbf{\bibinfo{volume}{516}}, \bibinfo{pages}{A66}
  (\bibinfo{year}{2010}), \eprint{1001.0551}.

\bibitem[{\citenamefont{{Trotta} et~al.}(2011)\citenamefont{{Trotta},
  {J{\'o}hannesson}, {Moskalenko}, {Porter}, {Ruiz de Austri}, and
  {Strong}}}]{2011ApJ...729..106T}
\bibinfo{author}{\bibfnamefont{R.}~\bibnamefont{{Trotta}}},
  \bibinfo{author}{\bibfnamefont{G.}~\bibnamefont{{J{\'o}hannesson}}},
  \bibinfo{author}{\bibfnamefont{I.~V.} \bibnamefont{{Moskalenko}}},
  \bibinfo{author}{\bibfnamefont{T.~A.} \bibnamefont{{Porter}}},
  \bibinfo{author}{\bibfnamefont{R.}~\bibnamefont{{Ruiz de Austri}}},
  \bibnamefont{and} \bibinfo{author}{\bibfnamefont{A.~W.}
  \bibnamefont{{Strong}}}, \bibinfo{journal}{\apj}
  \textbf{\bibinfo{volume}{729}}, \bibinfo{pages}{106} (\bibinfo{year}{2011}),
  \eprint{1011.0037}.

\bibitem[{\citenamefont{{Jin} et~al.}(2015)\citenamefont{{Jin}, {Wu}, and
  {Zhou}}}]{2015JCAP...09..049J}
\bibinfo{author}{\bibfnamefont{H.-B.} \bibnamefont{{Jin}}},
  \bibinfo{author}{\bibfnamefont{Y.-L.} \bibnamefont{{Wu}}}, \bibnamefont{and}
  \bibinfo{author}{\bibfnamefont{Y.-F.} \bibnamefont{{Zhou}}},
  \bibinfo{journal}{\jcap} \textbf{\bibinfo{volume}{9}}, \bibinfo{eid}{049}
  (\bibinfo{year}{2015}), \eprint{1410.0171}.

\bibitem[{\citenamefont{{J{\'o}hannesson}
  et~al.}(2016)\citenamefont{{J{\'o}hannesson}, {Ruiz de Austri}, {Vincent},
  {Moskalenko}, {Orlando}, {Porter}, {Strong}, {Trotta}, {Feroz}, {Graff}
  et~al.}}]{2016ApJ...824...16J}
\bibinfo{author}{\bibfnamefont{G.}~\bibnamefont{{J{\'o}hannesson}}},
  \bibnamefont{et~al.}, \bibinfo{journal}{\apj} \textbf{\bibinfo{volume}{824}},
  \bibinfo{eid}{16} (\bibinfo{year}{2016}), \eprint{1602.02243}.

\bibitem[{\citenamefont{{Feng} et~al.}(2016)\citenamefont{{Feng}, {Tomassetti},
  and {Oliva}}}]{2016PhRvD..94l3007F}
\bibinfo{author}{\bibfnamefont{J.}~\bibnamefont{{Feng}}},
  \bibinfo{author}{\bibfnamefont{N.}~\bibnamefont{{Tomassetti}}},
  \bibnamefont{and} \bibinfo{author}{\bibfnamefont{A.}~\bibnamefont{{Oliva}}},
  \bibinfo{journal}{\prd} \textbf{\bibinfo{volume}{94}}, \bibinfo{eid}{123007}
  (\bibinfo{year}{2016}), \eprint{1610.06182}.

\bibitem[{\citenamefont{{Korsmeier} and {Cuoco}}(2016)}]{2016PhRvD..94l3019K}
\bibinfo{author}{\bibfnamefont{M.}~\bibnamefont{{Korsmeier}}} \bibnamefont{and}
  \bibinfo{author}{\bibfnamefont{A.}~\bibnamefont{{Cuoco}}},
  \bibinfo{journal}{\prd} \textbf{\bibinfo{volume}{94}}, \bibinfo{eid}{123019}
  (\bibinfo{year}{2016}), \eprint{1607.06093}.

\bibitem[{\citenamefont{{Lewis} and {Bridle}}(2002)}]{2002PhRvD..66j3511L}
\bibinfo{author}{\bibfnamefont{A.}~\bibnamefont{{Lewis}}} \bibnamefont{and}
  \bibinfo{author}{\bibfnamefont{S.}~\bibnamefont{{Bridle}}},
  \bibinfo{journal}{\prd} \textbf{\bibinfo{volume}{66}},
  \bibinfo{pages}{103511} (\bibinfo{year}{2002}), \eprint{astro-ph/0205436}.

\bibitem[{\citenamefont{{Liu} et~al.}(2010)\citenamefont{{Liu}, {Yuan}, {Bi},
  {Li}, and {Zhang}}}]{2010PhRvD..81b3516L}
\bibinfo{author}{\bibfnamefont{J.}~\bibnamefont{{Liu}}},
  \bibinfo{author}{\bibfnamefont{Q.}~\bibnamefont{{Yuan}}},
  \bibinfo{author}{\bibfnamefont{X.~J.} \bibnamefont{{Bi}}},
  \bibinfo{author}{\bibfnamefont{H.}~\bibnamefont{{Li}}}, \bibnamefont{and}
  \bibinfo{author}{\bibfnamefont{X.~M.} \bibnamefont{{Zhang}}},
  \bibinfo{journal}{\prd} \textbf{\bibinfo{volume}{81}},
  \bibinfo{pages}{023516} (\bibinfo{year}{2010}), \eprint{0906.3858}.

\bibitem[{\citenamefont{{Liu} et~al.}(2012{\natexlab{a}})\citenamefont{{Liu},
  {Yuan}, {Bi}, {Li}, and {Zhang}}}]{2012IJMPA..2750024L}
\bibinfo{author}{\bibfnamefont{J.}~\bibnamefont{{Liu}}},
  \bibinfo{author}{\bibfnamefont{Q.}~\bibnamefont{{Yuan}}},
  \bibinfo{author}{\bibfnamefont{X.}~\bibnamefont{{Bi}}},
  \bibinfo{author}{\bibfnamefont{H.}~\bibnamefont{{Li}}}, \bibnamefont{and}
  \bibinfo{author}{\bibfnamefont{X.}~\bibnamefont{{Zhang}}},
  \bibinfo{journal}{International Journal of Modern Physics A}
  \textbf{\bibinfo{volume}{27}}, \bibinfo{eid}{1250024}
  (\bibinfo{year}{2012}{\natexlab{a}}), \eprint{0911.1002}.

\bibitem[{\citenamefont{{Liu} et~al.}(2012{\natexlab{b}})\citenamefont{{Liu},
  {Yuan}, {Bi}, {Li}, and {Zhang}}}]{2012PhRvD..85d3507L}
\bibinfo{author}{\bibfnamefont{J.}~\bibnamefont{{Liu}}},
  \bibinfo{author}{\bibfnamefont{Q.}~\bibnamefont{{Yuan}}},
  \bibinfo{author}{\bibfnamefont{X.-J.} \bibnamefont{{Bi}}},
  \bibinfo{author}{\bibfnamefont{H.}~\bibnamefont{{Li}}}, \bibnamefont{and}
  \bibinfo{author}{\bibfnamefont{X.}~\bibnamefont{{Zhang}}},
  \bibinfo{journal}{\prd} \textbf{\bibinfo{volume}{85}}, \bibinfo{eid}{043507}
  (\bibinfo{year}{2012}{\natexlab{b}}), \eprint{1106.3882}.

\bibitem[{\citenamefont{{Yuan} et~al.}(2015)\citenamefont{{Yuan}, {Bi}, {Chen},
  {Guo}, {Lin}, and {Zhang}}}]{2015APh....60....1Y}
\bibinfo{author}{\bibfnamefont{Q.}~\bibnamefont{{Yuan}}},
  \bibinfo{author}{\bibfnamefont{X.-J.} \bibnamefont{{Bi}}},
  \bibinfo{author}{\bibfnamefont{G.-M.} \bibnamefont{{Chen}}},
  \bibinfo{author}{\bibfnamefont{Y.-Q.} \bibnamefont{{Guo}}},
  \bibinfo{author}{\bibfnamefont{S.-J.} \bibnamefont{{Lin}}}, \bibnamefont{and}
  \bibinfo{author}{\bibfnamefont{X.}~\bibnamefont{{Zhang}}},
  \bibinfo{journal}{Astroparticle Physics} \textbf{\bibinfo{volume}{60}},
  \bibinfo{pages}{1} (\bibinfo{year}{2015}), \eprint{1304.1482}.

\bibitem[{\citenamefont{{Yuan} and {Bi}}(2013)}]{2013PhLB..727....1Y}
\bibinfo{author}{\bibfnamefont{Q.}~\bibnamefont{{Yuan}}} \bibnamefont{and}
  \bibinfo{author}{\bibfnamefont{X.-J.} \bibnamefont{{Bi}}},
  \bibinfo{journal}{Physics Letters B} \textbf{\bibinfo{volume}{727}},
  \bibinfo{pages}{1} (\bibinfo{year}{2013}), \eprint{1304.2687}.

\bibitem[{\citenamefont{{Yuan} and {Bi}}(2015)}]{2015JCAP...03..033Y}
\bibinfo{author}{\bibfnamefont{Q.}~\bibnamefont{{Yuan}}} \bibnamefont{and}
  \bibinfo{author}{\bibfnamefont{X.-J.} \bibnamefont{{Bi}}},
  \bibinfo{journal}{\jcap} \textbf{\bibinfo{volume}{3}}, \bibinfo{eid}{033}
  (\bibinfo{year}{2015}), \eprint{1408.2424}.

\bibitem[{\citenamefont{{Seo} and {Ptuskin}}(1994)}]{1994ApJ...431..705S}
\bibinfo{author}{\bibfnamefont{E.~S.} \bibnamefont{{Seo}}} \bibnamefont{and}
  \bibinfo{author}{\bibfnamefont{V.~S.} \bibnamefont{{Ptuskin}}},
  \bibinfo{journal}{\apj} \textbf{\bibinfo{volume}{431}}, \bibinfo{pages}{705}
  (\bibinfo{year}{1994}).

\bibitem[{\citenamefont{{Ackermann} et~al.}(2013)\citenamefont{{Ackermann},
  {Ajello}, {Allafort}, {Baldini}, {Ballet}, {Barbiellini}, {Baring},
  {Bastieri}, {Bechtol}, {Bellazzini} et~al.}}]{2013Sci...339..807A}
\bibinfo{author}{\bibfnamefont{M.}~\bibnamefont{{Ackermann}}},
  \bibnamefont{et~al.}, \bibinfo{journal}{Science}
  \textbf{\bibinfo{volume}{339}}, \bibinfo{pages}{807} (\bibinfo{year}{2013}),
  \eprint{1302.3307}.

\bibitem[{\citenamefont{{Panov} et~al.}(2007)\citenamefont{{Panov}, {Adams},
  {Ahn}, {Batkov}, {Bashindzhagyan}, {Watts}, {Wefel}, {Wu}, {Ganel}, {Guzik}
  et~al.}}]{2007BRASP..71..494P}
\bibinfo{author}{\bibfnamefont{A.~D.} \bibnamefont{{Panov}}},
  \bibnamefont{et~al.}, \bibinfo{journal}{Bulletin of the Russian Academy of
  Science, Phys.} \textbf{\bibinfo{volume}{71}}, \bibinfo{pages}{494}
  (\bibinfo{year}{2007}), \eprint{astro-ph/0612377}.

\bibitem[{\citenamefont{{Ahn} et~al.}(2010)\citenamefont{{Ahn}, {Allison},
  {Bagliesi}, {Beatty}, {Bigongiari}, {Childers}, {Conklin}, {Coutu},
  {DuVernois}, {Ganel} et~al.}}]{2010ApJ...714L..89A}
\bibinfo{author}{\bibfnamefont{H.~S.} \bibnamefont{{Ahn}}},
  \bibnamefont{et~al.}, \bibinfo{journal}{\apjl}
  \textbf{\bibinfo{volume}{714}}, \bibinfo{pages}{L89} (\bibinfo{year}{2010}),
  \eprint{1004.1123}.

\bibitem[{\citenamefont{{Adriani} et~al.}(2011)\citenamefont{{Adriani},
  {Barbarino}, {Bazilevskaya}, {Bellotti}, {Boezio}, {Bogomolov}, {Bonechi},
  {Bongi}, {Bonvicini}, {Borisov} et~al.}}]{2011Sci...332...69A}
\bibinfo{author}{\bibfnamefont{O.}~\bibnamefont{{Adriani}}},
  \bibnamefont{et~al.}, \bibinfo{journal}{Science}
  \textbf{\bibinfo{volume}{332}}, \bibinfo{pages}{69} (\bibinfo{year}{2011}),
  \eprint{1103.4055}.

\bibitem[{\citenamefont{{Aguilar}
  et~al.}(2015{\natexlab{a}})\citenamefont{{Aguilar}, {Aisa}, {Alpat},
  {Alvino}, {Ambrosi}, {Andeen}, {Arruda}, {Attig}, {Azzarello}, {Bachlechner}
  et~al.}}]{2015PhRvL.114q1103A}
\bibinfo{author}{\bibfnamefont{M.}~\bibnamefont{{Aguilar}}},
  \bibnamefont{et~al.}, \bibinfo{journal}{\prl} \textbf{\bibinfo{volume}{114}},
  \bibinfo{eid}{171103} (\bibinfo{year}{2015}{\natexlab{a}}).

\bibitem[{\citenamefont{{Aguilar}
  et~al.}(2015{\natexlab{b}})\citenamefont{{Aguilar}, {Aisa}, {Alpat},
  {Alvino}, {Ambrosi}, {Andeen}, {Arruda}, {Attig}, {Azzarello}, {Bachlechner}
  et~al.}}]{2015PhRvL.115u1101A}
\bibinfo{author}{\bibfnamefont{M.}~\bibnamefont{{Aguilar}}},
  \bibnamefont{et~al.}, \bibinfo{journal}{\prl} \textbf{\bibinfo{volume}{115}},
  \bibinfo{eid}{211101} (\bibinfo{year}{2015}{\natexlab{b}}).

\bibitem[{\citenamefont{{Jokipii}}(1976)}]{1976ApJ...208..900J}
\bibinfo{author}{\bibfnamefont{J.~R.} \bibnamefont{{Jokipii}}},
  \bibinfo{journal}{\apj} \textbf{\bibinfo{volume}{208}}, \bibinfo{pages}{900}
  (\bibinfo{year}{1976}).

\bibitem[{\citenamefont{{Engelmann} et~al.}(1990)\citenamefont{{Engelmann},
  {Ferrando}, {Soutoul}, {Goret}, and {Juliusson}}}]{1990A&A...233...96E}
\bibinfo{author}{\bibfnamefont{J.~J.} \bibnamefont{{Engelmann}}},
  \bibinfo{author}{\bibfnamefont{P.}~\bibnamefont{{Ferrando}}},
  \bibinfo{author}{\bibfnamefont{A.}~\bibnamefont{{Soutoul}}},
  \bibinfo{author}{\bibfnamefont{P.}~\bibnamefont{{Goret}}}, \bibnamefont{and}
  \bibinfo{author}{\bibfnamefont{E.}~\bibnamefont{{Juliusson}}},
  \bibinfo{journal}{\aap} \textbf{\bibinfo{volume}{233}}, \bibinfo{pages}{96}
  (\bibinfo{year}{1990}).

\bibitem[{\citenamefont{{Moskalenko} et~al.}(2002)\citenamefont{{Moskalenko},
  {Strong}, {Ormes}, and {Potgieter}}}]{2002ApJ...565..280M}
\bibinfo{author}{\bibfnamefont{I.~V.} \bibnamefont{{Moskalenko}}},
  \bibinfo{author}{\bibfnamefont{A.~W.} \bibnamefont{{Strong}}},
  \bibinfo{author}{\bibfnamefont{J.~F.} \bibnamefont{{Ormes}}},
  \bibnamefont{and} \bibinfo{author}{\bibfnamefont{M.~S.}
  \bibnamefont{{Potgieter}}}, \bibinfo{journal}{\apj}
  \textbf{\bibinfo{volume}{565}}, \bibinfo{pages}{280} (\bibinfo{year}{2002}),
  \eprint{astro-ph/0106567}.

\bibitem[{\citenamefont{{Ptuskin} et~al.}(2006)\citenamefont{{Ptuskin},
  {Moskalenko}, {Jones}, {Strong}, and {Zirakashvili}}}]{2006ApJ...642..902P}
\bibinfo{author}{\bibfnamefont{V.~S.} \bibnamefont{{Ptuskin}}},
  \bibinfo{author}{\bibfnamefont{I.~V.} \bibnamefont{{Moskalenko}}},
  \bibinfo{author}{\bibfnamefont{F.~C.} \bibnamefont{{Jones}}},
  \bibinfo{author}{\bibfnamefont{A.~W.} \bibnamefont{{Strong}}},
  \bibnamefont{and} \bibinfo{author}{\bibfnamefont{V.~N.}
  \bibnamefont{{Zirakashvili}}}, \bibinfo{journal}{\apj}
  \textbf{\bibinfo{volume}{642}}, \bibinfo{pages}{902} (\bibinfo{year}{2006}),
  \eprint{astro-ph/0510335}.

\bibitem[{\citenamefont{{Tomassetti}}(2012)}]{2012ApJ...752L..13T}
\bibinfo{author}{\bibfnamefont{N.}~\bibnamefont{{Tomassetti}}},
  \bibinfo{journal}{\apjl} \textbf{\bibinfo{volume}{752}}, \bibinfo{eid}{L13}
  (\bibinfo{year}{2012}), \eprint{1204.4492}.

\bibitem[{\citenamefont{{Guo} et~al.}(2016)\citenamefont{{Guo}, {Tian}, and
  {Jin}}}]{2016ApJ...819...54G}
\bibinfo{author}{\bibfnamefont{Y.-Q.} \bibnamefont{{Guo}}},
  \bibinfo{author}{\bibfnamefont{Z.}~\bibnamefont{{Tian}}}, \bibnamefont{and}
  \bibinfo{author}{\bibfnamefont{C.}~\bibnamefont{{Jin}}},
  \bibinfo{journal}{\apj} \textbf{\bibinfo{volume}{819}}, \bibinfo{eid}{54}
  (\bibinfo{year}{2016}).

\bibitem[{\citenamefont{{D'Angelo} et~al.}(2016)\citenamefont{{D'Angelo},
  {Blasi}, and {Amato}}}]{2016PhRvD..94h3003D}
\bibinfo{author}{\bibfnamefont{M.}~\bibnamefont{{D'Angelo}}},
  \bibinfo{author}{\bibfnamefont{P.}~\bibnamefont{{Blasi}}}, \bibnamefont{and}
  \bibinfo{author}{\bibfnamefont{E.}~\bibnamefont{{Amato}}},
  \bibinfo{journal}{\prd} \textbf{\bibinfo{volume}{94}}, \bibinfo{eid}{083003}
  (\bibinfo{year}{2016}), \eprint{1512.05000}.

\bibitem[{\citenamefont{{Neal}}(1993)}]{Neal1993}
\bibinfo{author}{\bibfnamefont{R.~M.} \bibnamefont{{Neal}}},
  \emph{\bibinfo{title}{{Probabilistic Inference Using Markov Chain Monte Carlo
  Methods}}} (\bibinfo{publisher}{{Department of Computer Science, University
  of Toronto}}, \bibinfo{year}{1993}).

\bibitem[{\citenamefont{{Gamerman}}(1997)}]{Gamerman1997}
\bibinfo{author}{\bibfnamefont{D.}~\bibnamefont{{Gamerman}}},
  \emph{\bibinfo{title}{{Markov Chain Monte Carlo: Stochastic Simulation for
  Bayesian Inference}}} (\bibinfo{publisher}{{Chapman and Hall, London}},
  \bibinfo{year}{1997}).

\bibitem[{\citenamefont{{Aguilar}
  et~al.}(2016{\natexlab{a}})\citenamefont{{Aguilar}, {Aisa}, {Alvino},
  {Ambrosi}, {Andeen}, {Arruda}, {Attig}, {Azzarello}, {Bachlechner}, {Barao}
  et~al.}}]{2016PhRvL.117w1102A}
\bibinfo{author}{\bibfnamefont{M.}~\bibnamefont{{Aguilar}}},
  \bibnamefont{et~al.}, \bibinfo{journal}{\prl} \textbf{\bibinfo{volume}{117}},
  \bibinfo{eid}{231102} (\bibinfo{year}{2016}{\natexlab{a}}).

\bibitem[{\citenamefont{{Connell}}(1998)}]{1998ApJ...501L..59C}
\bibinfo{author}{\bibfnamefont{J.~J.} \bibnamefont{{Connell}}},
  \bibinfo{journal}{\apjl} \textbf{\bibinfo{volume}{501}}, \bibinfo{pages}{L59}
  (\bibinfo{year}{1998}).

\bibitem[{\citenamefont{{Lukasiak}}(1999)}]{1999ICRC....3...41L}
\bibinfo{author}{\bibfnamefont{A.}~\bibnamefont{{Lukasiak}}}, in
  \emph{\bibinfo{booktitle}{International Cosmic Ray Conference}}
  (\bibinfo{year}{1999}), vol.~\bibinfo{volume}{3}, p.~\bibinfo{pages}{41}.

\bibitem[{\citenamefont{{Simpson} and
  {Garcia-Munoz}}(1988)}]{1988SSRv...46..205S}
\bibinfo{author}{\bibfnamefont{J.~A.} \bibnamefont{{Simpson}}}
  \bibnamefont{and}
  \bibinfo{author}{\bibfnamefont{M.}~\bibnamefont{{Garcia-Munoz}}},
  \bibinfo{journal}{Space Sci. Rev.} \textbf{\bibinfo{volume}{46}},
  \bibinfo{pages}{205} (\bibinfo{year}{1988}).

\bibitem[{\citenamefont{{Hams} et~al.}(2004)\citenamefont{{Hams}, {Barbier},
  {Bremerich}, {Christian}, {de Nolfo}, {Geier}, {G{\"o}bel}, {Gupta}, {Hof},
  {Menn} et~al.}}]{2004ApJ...611..892H}
\bibinfo{author}{\bibfnamefont{T.}~\bibnamefont{{Hams}}}, \bibnamefont{et~al.},
  \bibinfo{journal}{\apj} \textbf{\bibinfo{volume}{611}}, \bibinfo{pages}{892}
  (\bibinfo{year}{2004}).

\bibitem[{\citenamefont{{Adriani} et~al.}(2013)\citenamefont{{Adriani},
  {Barbarino}, {Bazilevskaya}, {Bellotti}, {Boezio}, {Bogomolov}, {Bongi},
  {Bonvicini}, {Borisov}, {Bottai} et~al.}}]{2013ApJ...765...91A}
\bibinfo{author}{\bibfnamefont{O.}~\bibnamefont{{Adriani}}},
  \bibnamefont{et~al.}, \bibinfo{journal}{\apj} \textbf{\bibinfo{volume}{765}},
  \bibinfo{eid}{91} (\bibinfo{year}{2013}), \eprint{1301.4108}.

\bibitem[{\citenamefont{{Gleeson} and {Axford}}(1968)}]{1968ApJ...154.1011G}
\bibinfo{author}{\bibfnamefont{L.~J.} \bibnamefont{{Gleeson}}}
  \bibnamefont{and} \bibinfo{author}{\bibfnamefont{W.~I.}
  \bibnamefont{{Axford}}}, \bibinfo{journal}{\apj}
  \textbf{\bibinfo{volume}{154}}, \bibinfo{pages}{1011} (\bibinfo{year}{1968}).

\bibitem[{\citenamefont{{Hathaway} et~al.}(1999)\citenamefont{{Hathaway},
  {Wilson}, and {Reichmann}}}]{1999JGR...10422375H}
\bibinfo{author}{\bibfnamefont{D.~H.} \bibnamefont{{Hathaway}}},
  \bibinfo{author}{\bibfnamefont{R.~M.} \bibnamefont{{Wilson}}},
  \bibnamefont{and} \bibinfo{author}{\bibfnamefont{E.~J.}
  \bibnamefont{{Reichmann}}}, \bibinfo{journal}{\jgr}
  \textbf{\bibinfo{volume}{104}}, \bibinfo{pages}{22375}
  (\bibinfo{year}{1999}).

\bibitem[{\citenamefont{{George} et~al.}(2009)\citenamefont{{George}, {Lave},
  {Wiedenbeck}, {Binns}, {Cummings}, {Davis}, {de Nolfo}, {Hink}, {Israel},
  {Leske} et~al.}}]{2009ApJ...698.1666G}
\bibinfo{author}{\bibfnamefont{J.~S.} \bibnamefont{{George}}},
  \bibnamefont{et~al.}, \bibinfo{journal}{\apj} \textbf{\bibinfo{volume}{698}},
  \bibinfo{pages}{1666} (\bibinfo{year}{2009}).

\bibitem[{\citenamefont{{Lin} et~al.}(2015)\citenamefont{{Lin}, {Yuan}, and
  {Bi}}}]{2015PhRvD..91f3508L}
\bibinfo{author}{\bibfnamefont{S.-J.} \bibnamefont{{Lin}}},
  \bibinfo{author}{\bibfnamefont{Q.}~\bibnamefont{{Yuan}}}, \bibnamefont{and}
  \bibinfo{author}{\bibfnamefont{X.-J.} \bibnamefont{{Bi}}},
  \bibinfo{journal}{\prd} \textbf{\bibinfo{volume}{91}}, \bibinfo{eid}{063508}
  (\bibinfo{year}{2015}), \eprint{1409.6248}.

\bibitem[{\citenamefont{{Di Bernardo} et~al.}(2013)\citenamefont{{Di Bernardo},
  {Evoli}, {Gaggero}, {Grasso}, and {Maccione}}}]{2013JCAP...03..036D}
\bibinfo{author}{\bibfnamefont{G.}~\bibnamefont{{Di Bernardo}}},
  \bibinfo{author}{\bibfnamefont{C.}~\bibnamefont{{Evoli}}},
  \bibinfo{author}{\bibfnamefont{D.}~\bibnamefont{{Gaggero}}},
  \bibinfo{author}{\bibfnamefont{D.}~\bibnamefont{{Grasso}}}, \bibnamefont{and}
  \bibinfo{author}{\bibfnamefont{L.}~\bibnamefont{{Maccione}}},
  \bibinfo{journal}{\jcap} \textbf{\bibinfo{volume}{3}}, \bibinfo{eid}{036}
  (\bibinfo{year}{2013}), \eprint{1210.4546}.

\bibitem[{\citenamefont{{Kolmogorov}}(1941)}]{1941DoSSR..30..301K}
\bibinfo{author}{\bibfnamefont{A.}~\bibnamefont{{Kolmogorov}}},
  \bibinfo{journal}{Akademiia Nauk SSSR Doklady} \textbf{\bibinfo{volume}{30}},
  \bibinfo{pages}{301} (\bibinfo{year}{1941}).

\bibitem[{\citenamefont{{Kraichnan}}(1965)}]{1965PhFl....8.1385K}
\bibinfo{author}{\bibfnamefont{R.~H.} \bibnamefont{{Kraichnan}}},
  \bibinfo{journal}{Physics of Fluids} \textbf{\bibinfo{volume}{8}},
  \bibinfo{pages}{1385} (\bibinfo{year}{1965}).

\bibitem[{\citenamefont{{Aguilar} et~al.}(2014)\citenamefont{{Aguilar}, {Aisa},
  {Alvino}, {Ambrosi}, {Andeen}, {Arruda}, {Attig}, {Azzarello}, {Bachlechner},
  {Barao} et~al.}}]{2014PhRvL.113l1102A}
\bibinfo{author}{\bibfnamefont{M.}~\bibnamefont{{Aguilar}}},
  \bibnamefont{et~al.}, \bibinfo{journal}{\prl} \textbf{\bibinfo{volume}{113}},
  \bibinfo{eid}{121102} (\bibinfo{year}{2014}).

\bibitem[{\citenamefont{{Shen}}(1970)}]{1970ApJ...162L.181S}
\bibinfo{author}{\bibfnamefont{C.~S.} \bibnamefont{{Shen}}},
  \bibinfo{journal}{\apjl} \textbf{\bibinfo{volume}{162}},
  \bibinfo{pages}{L181} (\bibinfo{year}{1970}).

\bibitem[{\citenamefont{{Harding} and {Ramaty}}(1987)}]{1987ICRC....2...92H}
\bibinfo{author}{\bibfnamefont{A.~K.} \bibnamefont{{Harding}}}
  \bibnamefont{and} \bibinfo{author}{\bibfnamefont{R.}~\bibnamefont{{Ramaty}}},
  \bibinfo{journal}{International Cosmic Ray Conference}
  \textbf{\bibinfo{volume}{2}}, \bibinfo{pages}{92} (\bibinfo{year}{1987}).

\bibitem[{\citenamefont{{Zhang} and {Cheng}}(2001)}]{2001A&A...368.1063Z}
\bibinfo{author}{\bibfnamefont{L.}~\bibnamefont{{Zhang}}} \bibnamefont{and}
  \bibinfo{author}{\bibfnamefont{K.~S.} \bibnamefont{{Cheng}}},
  \bibinfo{journal}{\aap} \textbf{\bibinfo{volume}{368}}, \bibinfo{pages}{1063}
  (\bibinfo{year}{2001}).

\bibitem[{\citenamefont{{Adriani} et~al.}(2010)\citenamefont{{Adriani},
  {Barbarino}, {Bazilevskaya}, {Bellotti}, {Boezio}, {Bogomolov}, {Bonechi},
  {Bongi}, {Bonvicini}, {Borisov} et~al.}}]{2010PhRvL.105l1101A}
\bibinfo{author}{\bibfnamefont{O.}~\bibnamefont{{Adriani}}},
  \bibnamefont{et~al.}, \bibinfo{journal}{\prl} \textbf{\bibinfo{volume}{105}},
  \bibinfo{pages}{121101} (\bibinfo{year}{2010}), \eprint{1007.0821}.

\bibitem[{\citenamefont{{Aguilar}
  et~al.}(2016{\natexlab{b}})\citenamefont{{Aguilar}, {Ali Cavasonza}, {Alpat},
  {Ambrosi}, {Arruda}, {Attig}, {Aupetit}, {Azzarello}, {Bachlechner}, {Barao}
  et~al.}}]{2016PhRvL.117i1103A}
\bibinfo{author}{\bibfnamefont{M.}~\bibnamefont{{Aguilar}}},
  \bibnamefont{et~al.}, \bibinfo{journal}{\prl} \textbf{\bibinfo{volume}{117}},
  \bibinfo{eid}{091103} (\bibinfo{year}{2016}{\natexlab{b}}).

\bibitem[{\citenamefont{{Huang} et~al.}(2016)\citenamefont{{Huang}, {Wei},
  {Wu}, {Zhang}, and {Zhou}}}]{2016arXiv161101983H}
\bibinfo{author}{\bibfnamefont{X.-J.} \bibnamefont{{Huang}}},
  \bibinfo{author}{\bibfnamefont{C.-C.} \bibnamefont{{Wei}}},
  \bibinfo{author}{\bibfnamefont{Y.-L.} \bibnamefont{{Wu}}},
  \bibinfo{author}{\bibfnamefont{W.-H.} \bibnamefont{{Zhang}}},
  \bibnamefont{and} \bibinfo{author}{\bibfnamefont{Y.-F.}
  \bibnamefont{{Zhou}}}, \bibinfo{journal}{ArXiv e-prints}
  (\bibinfo{year}{2016}), \eprint{1611.01983}.

\bibitem[{\citenamefont{{Li}}(2016)}]{2016arXiv161209501L}
\bibinfo{author}{\bibfnamefont{T.}~\bibnamefont{{Li}}}, \bibinfo{journal}{ArXiv
  e-prints}  (\bibinfo{year}{2016}), \eprint{1612.09501}.

\bibitem[{\citenamefont{{Feng} and {Zhang}}(2017)}]{2017arXiv170102263F}
\bibinfo{author}{\bibfnamefont{J.}~\bibnamefont{{Feng}}} \bibnamefont{and}
  \bibinfo{author}{\bibfnamefont{H.-H.} \bibnamefont{{Zhang}}},
  \bibinfo{journal}{ArXiv e-prints}  (\bibinfo{year}{2017}),
  \eprint{1701.02263}.

\bibitem[{\citenamefont{{Cholis} et~al.}(2017)\citenamefont{{Cholis}, {Hooper},
  and {Linden}}}]{2017arXiv170104406C}
\bibinfo{author}{\bibfnamefont{I.}~\bibnamefont{{Cholis}}},
  \bibinfo{author}{\bibfnamefont{D.}~\bibnamefont{{Hooper}}}, \bibnamefont{and}
  \bibinfo{author}{\bibfnamefont{T.}~\bibnamefont{{Linden}}},
  \bibinfo{journal}{ArXiv e-prints}  (\bibinfo{year}{2017}),
  \eprint{1701.04406}.

\bibitem[{\citenamefont{{AMS-02 collaboration}}(2016)}]{2016-AMS02-CERN}
\bibinfo{author}{\bibnamefont{{AMS-02 collaboration}}}, in
  \emph{\bibinfo{booktitle}{AMS Five Years Data Release}}
  (\bibinfo{publisher}{http://www.ams02.org/}, \bibinfo{year}{2016}).

\bibitem[{\citenamefont{{Kamae} et~al.}(2006)\citenamefont{{Kamae}, {Karlsson},
  {Mizuno}, {Abe}, and {Koi}}}]{2006ApJ...647..692K}
\bibinfo{author}{\bibfnamefont{T.}~\bibnamefont{{Kamae}}},
  \bibinfo{author}{\bibfnamefont{N.}~\bibnamefont{{Karlsson}}},
  \bibinfo{author}{\bibfnamefont{T.}~\bibnamefont{{Mizuno}}},
  \bibinfo{author}{\bibfnamefont{T.}~\bibnamefont{{Abe}}}, \bibnamefont{and}
  \bibinfo{author}{\bibfnamefont{T.}~\bibnamefont{{Koi}}},
  \bibinfo{journal}{\apj} \textbf{\bibinfo{volume}{647}}, \bibinfo{pages}{692}
  (\bibinfo{year}{2006}), \eprint{astro-ph/0605581}.

\bibitem[{\citenamefont{{Delahaye} et~al.}(2009)\citenamefont{{Delahaye},
  {Lineros}, {Donato}, {Fornengo}, {Lavalle}, {Salati}, and
  {Taillet}}}]{2009A&A...501..821D}
\bibinfo{author}{\bibfnamefont{T.}~\bibnamefont{{Delahaye}}},
  \bibinfo{author}{\bibfnamefont{R.}~\bibnamefont{{Lineros}}},
  \bibinfo{author}{\bibfnamefont{F.}~\bibnamefont{{Donato}}},
  \bibinfo{author}{\bibfnamefont{N.}~\bibnamefont{{Fornengo}}},
  \bibinfo{author}{\bibfnamefont{J.}~\bibnamefont{{Lavalle}}},
  \bibinfo{author}{\bibfnamefont{P.}~\bibnamefont{{Salati}}}, \bibnamefont{and}
  \bibinfo{author}{\bibfnamefont{R.}~\bibnamefont{{Taillet}}},
  \bibinfo{journal}{\aap} \textbf{\bibinfo{volume}{501}}, \bibinfo{pages}{821}
  (\bibinfo{year}{2009}), \eprint{0809.5268}.

\bibitem[{\citenamefont{{Pohl} and {Esposito}}(1998)}]{1998ApJ...507..327P}
\bibinfo{author}{\bibfnamefont{M.}~\bibnamefont{{Pohl}}} \bibnamefont{and}
  \bibinfo{author}{\bibfnamefont{J.~A.} \bibnamefont{{Esposito}}},
  \bibinfo{journal}{\apj} \textbf{\bibinfo{volume}{507}}, \bibinfo{pages}{327}
  (\bibinfo{year}{1998}), \eprint{astro-ph/9806160}.

\bibitem[{\citenamefont{{Clem} et~al.}(1996)\citenamefont{{Clem}, {Clements},
  {Esposito}, {Evenson}, {Huber}, {L'Heureux}, {Meyer}, and
  {Constantin}}}]{1996ApJ...464..507C}
\bibinfo{author}{\bibfnamefont{J.~M.} \bibnamefont{{Clem}}},
  \bibinfo{author}{\bibfnamefont{D.~P.} \bibnamefont{{Clements}}},
  \bibinfo{author}{\bibfnamefont{J.}~\bibnamefont{{Esposito}}},
  \bibinfo{author}{\bibfnamefont{P.}~\bibnamefont{{Evenson}}},
  \bibinfo{author}{\bibfnamefont{D.}~\bibnamefont{{Huber}}},
  \bibinfo{author}{\bibfnamefont{J.}~\bibnamefont{{L'Heureux}}},
  \bibinfo{author}{\bibfnamefont{P.}~\bibnamefont{{Meyer}}}, \bibnamefont{and}
  \bibinfo{author}{\bibfnamefont{C.}~\bibnamefont{{Constantin}}},
  \bibinfo{journal}{\apj} \textbf{\bibinfo{volume}{464}}, \bibinfo{pages}{507}
  (\bibinfo{year}{1996}).

\bibitem[{\citenamefont{{Della Torre} et~al.}(2012)\citenamefont{{Della Torre},
  {Bobik}, {Boschini}, {Consolandi}, {Gervasi}, {Grandi}, {Kudela}, {Pensotti},
  {Rancoita}, {Rozza} et~al.}}]{2012AdSpR..49.1587D}
\bibinfo{author}{\bibfnamefont{S.}~\bibnamefont{{Della Torre}}},
  \bibnamefont{et~al.}, \bibinfo{journal}{Advances in Space Research}
  \textbf{\bibinfo{volume}{49}}, \bibinfo{pages}{1587} (\bibinfo{year}{2012}).

\bibitem[{\citenamefont{{Maccione}}(2013)}]{2013PhRvL.110h1101M}
\bibinfo{author}{\bibfnamefont{L.}~\bibnamefont{{Maccione}}},
  \bibinfo{journal}{\prl} \textbf{\bibinfo{volume}{110}}, \bibinfo{eid}{081101}
  (\bibinfo{year}{2013}), \eprint{1211.6905}.

\bibitem[{\citenamefont{{Potgieter} et~al.}(2014)\citenamefont{{Potgieter},
  {Vos}, {Boezio}, {De Simone}, {Di Felice}, and
  {Formato}}}]{2014SoPh..289..391P}
\bibinfo{author}{\bibfnamefont{M.~S.} \bibnamefont{{Potgieter}}},
  \bibinfo{author}{\bibfnamefont{E.~E.} \bibnamefont{{Vos}}},
  \bibinfo{author}{\bibfnamefont{M.}~\bibnamefont{{Boezio}}},
  \bibinfo{author}{\bibfnamefont{N.}~\bibnamefont{{De Simone}}},
  \bibinfo{author}{\bibfnamefont{V.}~\bibnamefont{{Di Felice}}},
  \bibnamefont{and}
  \bibinfo{author}{\bibfnamefont{V.}~\bibnamefont{{Formato}}},
  \bibinfo{journal}{Solar Physics} \textbf{\bibinfo{volume}{289}},
  \bibinfo{pages}{391} (\bibinfo{year}{2014}), \eprint{1302.1284}.

\bibitem[{\citenamefont{{Kappl}}(2016)}]{2016CoPhC.207..386K}
\bibinfo{author}{\bibfnamefont{R.}~\bibnamefont{{Kappl}}},
  \bibinfo{journal}{Computer Physics Communications}
  \textbf{\bibinfo{volume}{207}}, \bibinfo{pages}{386} (\bibinfo{year}{2016}),
  \eprint{1511.07875}.

\bibitem[{\citenamefont{{Stone} et~al.}(2013)\citenamefont{{Stone}, {Cummings},
  {McDonald}, {Heikkila}, {Lal}, and {Webber}}}]{2013Sci...341..150S}
\bibinfo{author}{\bibfnamefont{E.~C.} \bibnamefont{{Stone}}},
  \bibinfo{author}{\bibfnamefont{A.~C.} \bibnamefont{{Cummings}}},
  \bibinfo{author}{\bibfnamefont{F.~B.} \bibnamefont{{McDonald}}},
  \bibinfo{author}{\bibfnamefont{B.~C.} \bibnamefont{{Heikkila}}},
  \bibinfo{author}{\bibfnamefont{N.}~\bibnamefont{{Lal}}}, \bibnamefont{and}
  \bibinfo{author}{\bibfnamefont{W.~R.} \bibnamefont{{Webber}}},
  \bibinfo{journal}{Science} \textbf{\bibinfo{volume}{341}},
  \bibinfo{pages}{150} (\bibinfo{year}{2013}).

\bibitem[{\citenamefont{{Cummings} et~al.}(2016)\citenamefont{{Cummings},
  {Stone}, {Heikkila}, {Lal}, {Webber}, {J{\'o}hannesson}, {Moskalenko},
  {Orlando}, and {Porter}}}]{2016ApJ...831...18C}
\bibinfo{author}{\bibfnamefont{A.~C.} \bibnamefont{{Cummings}}},
  \bibinfo{author}{\bibfnamefont{E.~C.} \bibnamefont{{Stone}}},
  \bibinfo{author}{\bibfnamefont{B.~C.} \bibnamefont{{Heikkila}}},
  \bibinfo{author}{\bibfnamefont{N.}~\bibnamefont{{Lal}}},
  \bibinfo{author}{\bibfnamefont{W.~R.} \bibnamefont{{Webber}}},
  \bibinfo{author}{\bibfnamefont{G.}~\bibnamefont{{J{\'o}hannesson}}},
  \bibinfo{author}{\bibfnamefont{I.~V.} \bibnamefont{{Moskalenko}}},
  \bibinfo{author}{\bibfnamefont{E.}~\bibnamefont{{Orlando}}},
  \bibnamefont{and} \bibinfo{author}{\bibfnamefont{T.~A.}
  \bibnamefont{{Porter}}}, \bibinfo{journal}{\apj}
  \textbf{\bibinfo{volume}{831}}, \bibinfo{eid}{18} (\bibinfo{year}{2016}).

\bibitem[{\citenamefont{{Moskalenko} et~al.}(2003)\citenamefont{{Moskalenko},
  {Strong}, {Mashnik}, and {Ormes}}}]{2003ApJ...586.1050M}
\bibinfo{author}{\bibfnamefont{I.~V.} \bibnamefont{{Moskalenko}}},
  \bibinfo{author}{\bibfnamefont{A.~W.} \bibnamefont{{Strong}}},
  \bibinfo{author}{\bibfnamefont{S.~G.} \bibnamefont{{Mashnik}}},
  \bibnamefont{and} \bibinfo{author}{\bibfnamefont{J.~F.}
  \bibnamefont{{Ormes}}}, \bibinfo{journal}{\apj}
  \textbf{\bibinfo{volume}{586}}, \bibinfo{pages}{1050} (\bibinfo{year}{2003}),
  \eprint{astro-ph/0210480}.

\bibitem[{\citenamefont{{Hooper} et~al.}(2015)\citenamefont{{Hooper}, {Linden},
  and {Mertsch}}}]{2015JCAP...03..021H}
\bibinfo{author}{\bibfnamefont{D.}~\bibnamefont{{Hooper}}},
  \bibinfo{author}{\bibfnamefont{T.}~\bibnamefont{{Linden}}}, \bibnamefont{and}
  \bibinfo{author}{\bibfnamefont{P.}~\bibnamefont{{Mertsch}}},
  \bibinfo{journal}{\jcap} \textbf{\bibinfo{volume}{3}}, \bibinfo{eid}{021}
  (\bibinfo{year}{2015}), \eprint{1410.1527}.

\bibitem[{\citenamefont{{Cui} et~al.}(2016)\citenamefont{{Cui}, {Yuan}, {Sming
  Tsai}, and {Fan}}}]{2016arXiv161003840C}
\bibinfo{author}{\bibfnamefont{M.-Y.} \bibnamefont{{Cui}}},
  \bibinfo{author}{\bibfnamefont{Q.}~\bibnamefont{{Yuan}}},
  \bibinfo{author}{\bibfnamefont{Y.-L.} \bibnamefont{{Sming Tsai}}},
  \bibnamefont{and} \bibinfo{author}{\bibfnamefont{Y.-Z.} \bibnamefont{{Fan}}},
  \bibinfo{journal}{ArXiv e-prints}  (\bibinfo{year}{2016}),
  \eprint{1610.03840}.

\bibitem[{\citenamefont{{Cuoco} et~al.}(2016)\citenamefont{{Cuoco}, {Kramer},
  and {Korsmeier}}}]{2016arXiv161003071C}
\bibinfo{author}{\bibfnamefont{A.}~\bibnamefont{{Cuoco}}},
  \bibinfo{author}{\bibfnamefont{M.}~\bibnamefont{{Kramer}}}, \bibnamefont{and}
  \bibinfo{author}{\bibfnamefont{M.}~\bibnamefont{{Korsmeier}}},
  \bibinfo{journal}{ArXiv e-prints}  (\bibinfo{year}{2016}),
  \eprint{1610.03071}.

\bibitem[{\citenamefont{{Donato} et~al.}(2001)\citenamefont{{Donato}, {Maurin},
  {Salati}, {Barrau}, {Boudoul}, and {Taillet}}}]{2001ApJ...563..172D}
\bibinfo{author}{\bibfnamefont{F.}~\bibnamefont{{Donato}}},
  \bibinfo{author}{\bibfnamefont{D.}~\bibnamefont{{Maurin}}},
  \bibinfo{author}{\bibfnamefont{P.}~\bibnamefont{{Salati}}},
  \bibinfo{author}{\bibfnamefont{A.}~\bibnamefont{{Barrau}}},
  \bibinfo{author}{\bibfnamefont{G.}~\bibnamefont{{Boudoul}}},
  \bibnamefont{and}
  \bibinfo{author}{\bibfnamefont{R.}~\bibnamefont{{Taillet}}},
  \bibinfo{journal}{\apj} \textbf{\bibinfo{volume}{563}}, \bibinfo{pages}{172}
  (\bibinfo{year}{2001}).

\bibitem[{\citenamefont{{Giesen} et~al.}(2015)\citenamefont{{Giesen},
  {Boudaud}, {G{\'e}nolini}, {Poulin}, {Cirelli}, {Salati}, and
  {Serpico}}}]{2015JCAP...09..023G}
\bibinfo{author}{\bibfnamefont{G.}~\bibnamefont{{Giesen}}},
  \bibinfo{author}{\bibfnamefont{M.}~\bibnamefont{{Boudaud}}},
  \bibinfo{author}{\bibfnamefont{Y.}~\bibnamefont{{G{\'e}nolini}}},
  \bibinfo{author}{\bibfnamefont{V.}~\bibnamefont{{Poulin}}},
  \bibinfo{author}{\bibfnamefont{M.}~\bibnamefont{{Cirelli}}},
  \bibinfo{author}{\bibfnamefont{P.}~\bibnamefont{{Salati}}}, \bibnamefont{and}
  \bibinfo{author}{\bibfnamefont{P.~D.} \bibnamefont{{Serpico}}},
  \bibinfo{journal}{\jcap} \textbf{\bibinfo{volume}{9}}, \bibinfo{eid}{023}
  (\bibinfo{year}{2015}), \eprint{1504.04276}.

\bibitem[{\citenamefont{{Lin} et~al.}(2016)\citenamefont{{Lin}, {Bi}, {Feng},
  {Yin}, and {Yu}}}]{2016arXiv161204001L}
\bibinfo{author}{\bibfnamefont{S.-J.} \bibnamefont{{Lin}}},
  \bibinfo{author}{\bibfnamefont{X.-J.} \bibnamefont{{Bi}}},
  \bibinfo{author}{\bibfnamefont{J.}~\bibnamefont{{Feng}}},
  \bibinfo{author}{\bibfnamefont{P.-F.} \bibnamefont{{Yin}}}, \bibnamefont{and}
  \bibinfo{author}{\bibfnamefont{Z.-H.} \bibnamefont{{Yu}}},
  \bibinfo{journal}{ArXiv e-prints}  (\bibinfo{year}{2016}),
  \eprint{1612.04001}.

\end{thebibliography}

\end{document}